\documentclass[
preprintnumbers, 
showpacs, 
amsmath, 
showkeys, 
amssymb, 
aps, 
superscriptaddress, 
twocolumn, 
longbibliography, 
letterpaper,
nofootinbib,
10pt
]{revtex4-2}
\usepackage{lipsum, babel}
\usepackage{color}
\usepackage{hyperref}
\usepackage{bm}
\usepackage{dcolumn} 
\usepackage{booktabs} 
\usepackage{amsfonts}
\usepackage{mathrsfs}
\usepackage{graphicx}
\usepackage{amsmath}
\usepackage{float}
\usepackage{braket}
\usepackage{natbib}
\usepackage{orcidlink}

     \hypersetup{
    colorlinks=true,
    linkcolor=red,
    citecolor=blue,
}

\begin{document}

\newcommand{\sgn}{\operatorname{sgn}}
\newcommand{\hhat}[1]{\hat {\hat{#1}}}
\newcommand{\pslash}[1]{#1\llap{\sl/}}
\newcommand{\kslash}[1]{\rlap{\sl/}#1}
\newcommand{\lab}[1]{}
\newcommand{\sto}[1]{\begin{center} \textit{#1} \end{center}}
\newcommand{\rf}[1]{{\color{blue}[\textit{#1}]}}
\newcommand{\eml}[1]{#1}
\newcommand{\el}[1]{\label{#1}}
\newcommand{\er}[1]{Eq.\eqref{#1}}
\newcommand{\df}[1]{\textbf{#1}}
\newcommand{\mdf}[1]{\pmb{#1}}
\newcommand{\ft}[1]{\footnote{#1}}
\newcommand{\n}[1]{$#1$}
\newcommand{\fals}[1]{$^\times$ #1}
\newcommand{\new}{{\color{red}$^{NEW}$ }}
\newcommand{\ci}[1]{}
\newcommand{\de}[1]{{\color{green}\underline{#1}}}
\newcommand{\ke}{\rangle}
\newcommand{\br}{\langle}
\newcommand{\lb}{\left(}
\newcommand{\rb}{\right)}
\newcommand{\lbk}{\left[}
\newcommand{\rbk}{\right]}
\newcommand{\blb}{\Big(}
\newcommand{\brb}{\Big)}
\newcommand{\nn}{\nonumber \\}
\newcommand{\p}{\partial}
\newcommand{\pd}[1]{\frac {\partial} {\partial #1}}
\newcommand{\cd}{\nabla}
\newcommand{\cc}{$>$}
\newcommand{\bqa}{\begin{eqnarray}}
\newcommand{\eqa}{\end{eqnarray}}
\newcommand{\bqe}{\begin{equation}}
\newcommand{\eqe}{\end{equation}}
\newcommand{\bay}[1]{\left(\begin{array}{#1}}
\newcommand{\eay}{\end{array}\right)}
\newcommand{\eg}{\textit{e.g.} }
\newcommand{\ie}{\textit{i.e.}, }
\newcommand{\iv}[1]{{#1}^{-1}}
\newcommand{\st}[1]{|#1\ke}
\newcommand{\at}[1]{{\Big|}_{#1}}
\newcommand{\zt}[1]{\texttt{#1}}
\newcommand{\non}{\nonumber}
\newcommand{\m}{\mu}
\def\xa{{m}}
\def\xA{{m}}
\def\xb{{\beta}}
\def\xB{{\Beta}}
\def\xd{{\delta}}
\def\xD{{\Delta}}
\def\xe{{\epsilon}}
\def\xE{{\Epsilon}}
\def\xve{{\varepsilon}}
\def\xg{{\gamma}}
\def\xG{{\Gamma}}
\def\xk{{\kappa}}
\def\xK{{\Kappa}}
\def\xl{{\lambda}}
\def\xL{{\Lambda}}
\def\xo{{\omega}}
\def\xO{{\Omega}}
\def\xvp{{\varphi}}
\def\xs{{\sigma}}
\def\xS{{\Sigma}}
\def\xt{{\theta}}
\def\xvt{{\vartheta}}
\def\xT{{\Theta}}
\def \Tr {{\rm Tr}}
\def\CA{{\cal A}}
\def\CC{{\cal C}}
\def\CD{{\cal D}}
\def\CE{{\cal E}}
\def\CF{{\cal F}}
\def\CH{{\cal H}}
\def\CJ{{\cal J}}
\def\CK{{\cal K}}
\def\CL{{\cal L}}
\def\CM{{\cal M}}
\def\CN{{\cal N}}
\def\CO{{\cal O}}
\def\CP{{\cal P}}
\def\CQ{{\cal Q}}
\def\CR{{\cal R}}
\def\CS{{\cal S}}
\def\CT{{\cal T}}
\def\CV{{\cal V}}
\def\CW{{\cal W}}
\def\CY{{\cal Y}}
\def\BC{\mathbb{C}}
\def\BR{\mathbb{R}}
\def\BZ{\mathbb{Z}}
\def\sA{\mathscr{A}}
\def\sB{\mathscr{B}}
\def\sF{\mathscr{F}}
\def\sG{\mathscr{G}}
\def\sH{\mathscr{H}}
\def\sJ{\mathscr{J}}
\def\sL{\mathscr{L}}
\def\sM{\mathscr{M}}
\def\sN{\mathscr{N}}
\def\sO{\mathscr{O}}
\def\sP{\mathscr{P}}
\def\sR{\mathscr{R}}
\def\sQ{\mathscr{Q}}
\def\sS{\mathscr{S}}
\def\sX{\mathscr{X}}



\author{Wei Kou\orcidlink{0000-0002-4152-2150}}
\email{kouwei@impcas.ac.cn}

\affiliation{Institute of Modern Physics, Chinese Academy of Sciences, Lanzhou 730000, China}
\affiliation{School of Nuclear Science and Technology, University of Chinese Academy of Sciences, Beijing 100049, China}
\author{Xiaoxuan Lin}
\affiliation{College of Physics Science and Technology, Hebei University, Baoding 071002, China}
\affiliation{Institute of Modern Physics, Chinese Academy of Sciences, Lanzhou 730000, China}

\author{Bing'ang Guo}

\affiliation{Institute of Modern Physics, Chinese Academy of Sciences, Lanzhou 730000, China}
\affiliation{School of Nuclear Science and Technology, University of Chinese Academy of Sciences, Beijing 100049, China}

\author{Xurong Chen}
\email{xchen@impcas.ac.cn (Corresponding Author)}
\affiliation{Institute of Modern Physics, Chinese Academy of Sciences, Lanzhou 730000, China}
\affiliation{School of Nuclear Science and Technology, University of Chinese Academy of Sciences, Beijing 100049, China}
\affiliation{Southern Center for Nuclear Science Theory (SCNT), Institute of Modern Physics, Chinese Academy of Sciences, Huizhou 516000, Guangdong Province, China}

\title{Physics-Informed Neural Network Approach to Quark–Antiquark Color Flux Tube}

\begin{abstract}

We introduce a physics-informed neural network (PINNs) framework for modelling the spatial distribution of chromodynamic fields induced by quark–antiquark pairs, based on lattice Monte Carlo simulations. In contrast to conventional neural networks, PINNs incorporate physical laws—expressed here as differential equations governing type-II superconductivity—directly into the training objective. By embedding these equations into the loss function, we guide the network to learn physically consistent solutions. Adopting an inverse problem approach, we extract the parameters of the superconducting equations from lattice QCD data and subsequently solve them. To accommodate physical boundary conditions, we recast the system into an integro-differential form and extend the analysis within the fractional PINNs framework. The accuracy of the reconstructed field distribution is assessed via relative $L_2$-error norms. We further extract physical observables such as the string tension and the mean width of the flux tube, offering quantitative insight into the confinement mechanism. This method enables the reconstruction of colour field profiles as functions of quark–antiquark separation without recourse to predefined parametric models. Our results illuminate aspects of the dual Meissner effect and highlight the promise of data-driven strategies in addressing non-perturbative challenges in quantum chromodynamics.
\end{abstract}


\maketitle

\section{Introduction}
\label{sec:introduction}
Quantum Chromodynamics (QCD), as the fundamental theory describing the strong interaction between quarks and gluons, has given rise to a comprehensive framework for hadronic physics through the study of its essential features—namely, asymptotic freedom and spontaneous chiral symmetry breaking. Due to confinement, the constituent degrees of freedom of hadrons—quarks and gluons—cannot be isolated, and confinement itself is an intrinsically non-perturbative phenomenon of QCD at low energies. Consequently, the theoretical treatment of QCD in this regime often relies heavily on phenomenological inputs. Fortunately, QCD is an experimentally grounded theory: extensive data have been accumulated over decades from large-scale collider and fixed-target experiments involving strong interactions. Moreover, advancements in computational hardware and algorithms have enabled lattice Monte Carlo simulations to provide initial yet valuable insights and measurements concerning non-perturbative phenomena in QCD.

In high-energy physics, the problem of confinement has remained unresolved for a long time, particularly within the framework of quantum field theory, where multiple theoretical approaches have been proposed. The dual superconductivity picture is one such hypothesis for the confinement mechanism \cite{Nambu:1974zg}. In this scenario, the vacuum may undergo magnetic monopole condensation, resulting in the formation of confining Abelian electric flux tubes that connect quarks and antiquarks. These flux tubes are the dual analogs of Abrikosov-Nielsen-Olesen (ANO) vortices in type-II superconductors \cite{Abrikosov:1956sx,Abrikosov:1957wnz,Nielsen:1973cs}. Detailed discussions on the dual superconductivity mechanism can be found in the review \cite{Kondo:2014sta}. Lattice QCD simulations have long been employed as one of the primary approaches for obtaining the chromoelectric field or the profile of the flux tube \cite{Singh:1993jj,Schilling:1998gz,Chernodub:2000rg,Chernodub:2005gz,Cea:2012qw}. It is worth noting that these studies largely rely on the choice of a specific gauge condition \cite{Suzuki:2009xy}, namely the so-called ``maximal Abelian projection”. Among numerous lattice simulation studies, the Clem ansatz \cite{clem1975simple} is frequently employed to describe the short-distance behavior of flux tubes. The Clem ansatz is intended to hypothesize or provide an approximate analytical expression for the transverse profile of the flux tube. More specifically, it introduces a simplified ordinary differential equation (ODE) that captures the spatial variation of the magnetic field or magnetic potential within the flux tube. This ODE is governed by the Ginzburg–Landau parameter, and under specific boundary conditions, its solutions are expressed as combinations of Bessel functions. Such solutions are often used to fit the chromoelectric field distributions obtained from lattice simulations.

Many researchers aspire to address such problems using a forward-thinking approach—for instance, by directly solving the second-type superconducting ODE associated with the Clem ansatz, and subsequently analyzing the parameterization of the analytical solution based on lattice simulation data. While studying the solutions of such equations is inherently meaningful, the high complexity of many problems often precludes the availability of explicit analytical expressions, thus hindering forward approaches.
Conversely, in QCD research, one frequently encounters inverse problems—namely, deducing physical laws from experimental observations. These problems are inherently challenging and demand sophisticated solution strategies. Various methods exist for inferring model parameters from experimental or simulated data, and recent advances in data-driven approaches have introduced increasingly modern techniques.
Machine learning (ML), a modern branch of artificial intelligence (AI), is gaining traction as a powerful tool in QCD studies and applications \cite{Larkoski:2017jix,Guest:2018yhq,Radovic:2018dip,Albertsson:2018maf,Carleo:2019ptp,Bourilkov:2019yoi,Schwartz:2021ftp,Karagiorgi:2021ngt,Boehnlein:2021eym,Shanahan:2022ifi,Yang:2022yfr,Li:2022ozl,He:2023zin,Zhou:2023pti,Zhou:2023tvv,Pang:2024kid,Ma:2023zfj,Luo:2024iwf,Chen:2024epd,OmanaKuttan:2023bnb,Shi:2022vfr,Shi:2022fei,Shi:2021qri,Mansouri:2024uwc,Chen:2024mmd,Chen:2024ckb,Wang:2023poi,Bento:2025agw} (see also the review \cite{Aarts:2025gyp} ). ML provides a data-driven framework for uncovering the behavior of complex systems. In particular, by training deep neural networks (DNNs) on empirical data, one can infer model parameters and associated physical properties. Furthermore, embedding physical knowledge—such as PDEs or ODEs and their symmetries—into DNNs can enhance the fidelity of the learning outcomes. This class of methods is known as physics-informed neural networks (PINNs) \cite{raissi2019physics}.

Physical equations are often employed as prior conditions for describing a system, with their associated laws and symmetries providing constraints that ensure the solutions obtained during neural network training remain physically consistent. PINNs offer a framework that leverages automatic differentiation (AD) \cite{Baydin:2015tfa} to handle PDEs or ODEs. By incorporating auxiliary loss terms corresponding to the governing equations, PINNs enable supervised learning constrained by physical laws, allowing the solution of both linear and nonlinear problems, such as the Navier–Stokes equation in fluid dynamics \cite{amalinadhi2022physics,eivazi2022physics}.
In this work, we introduce the PINNs framework into the study of chromoelectric field distributions. By analyzing color flux tube data generated from lattice simulations of quark–antiquark excitations at varying distances \cite{Baker:2019gsi}, we aim to solve the simplified Ginzburg–Landau equation embedded in the Clem ansatz \cite{clem1975simple}. Without imposing explicit boundary conditions, and following a purely data-driven approach, we successfully infer both the field profiles and the governing parameters corresponding to different quark separations. The PINNs functionality implemented in the DeepXDE \cite{lu2021deepxde} software package is used to perform the analysis. The structure of this paper is organized as follows: in Sec. \ref{sec:pde}, we introduce the chromoelectric field distributions and outline relevant details of the lattice simulations. Sec. \ref{sec:PINN} provides a review of the key concepts underlying PINNs, with a particular focus on the formulation of PDEs and the corresponding loss functions. The Sec. \ref{sec:results} presents our results and discussion. Finally, we offer a summary and concluding remarks.

\section{Simple model in a type II superconductor}
\label{sec:pde}
In the dual superconductivity framework, the chromoelectric field is confined into a flux tube via the dual Meissner effect, analogous to the characteristics of an ANO vortex. Type-I and type-II superconductors exhibit attractive and repulsive forces, respectively, along the vortex axis. From the superconducting analogy, the type of dual superconductivity characterizes the vacuum of a Yang–Mills theory or QCD in which quark confinement occurs.
In Clem’s work \cite{clem1975simple}, it was pointed out that the penetration depth and coherence length calculated within the London model diverge along the vortex axis, due to the model’s neglect of the vanishing order parameter at the core. To address this issue, Clem proposed a variational model of a normalized order parameter for an isolated vortex and derived simple analytical expressions for the magnetic field and supercurrent density that satisfy both Ampère’s law and the Ginzburg–Landau equation. It is worth noting that this simplified model provides a remarkably good description of the chromoelectric field profiles obtained from lattice simulations \cite{Cardaci:2010tb,Cea:2012qw,Cea:2013oba,Cea:2014hma,Cea:2014uja}, offering a practical approach for determining superconducting parameters in studies of the chromoelectric field distribution and dual superconductivity on the lattice.

In Ref. \cite{clem1975simple}, the author presented a simple variational model for the magnitude of the normalized order paraemer of an isolated vortex: $f=x_t/R$, where $x_t$ is the radial coordinate \footnote{In this work, we adopt cylindrical coordinates. Assuming that the chromoelectric field distribution is independent of the $z$-coordinate, it can then be treated as a function of the transverse spatial coordinate.}, $R=\sqrt{x_t^2+\xi_v^2}$, and $\xi_v$ is a variational core radius parameter. In the cylindrical coordinate framework, the magnetic flux density $\mathbf{b}=\hat{z}b_z(x_t)$ is oriented along the $\hat{z}$ direction and depends on the radial coordinate $x_t$. Under this configuration, the supercurrent density can be expressed as $\mathbf{j} = \hat{\varphi} j_\varphi(x_t)$. Introducing a vector potential $\mathbf{a} = \hat{\varphi}a_\varphi(x_t)$, which satisfies the Coulomb gauge and vanishes at the vortex core, the second Ginzburg–Landau equation can be written as:
\begin{equation}
	j_\varphi=-\left(\frac{1}{4\pi\lambda^2}\right)\left(a_\varphi-\frac{\phi}{2\pi x_t}\right)f^2,
	\label{eq:G-L-E}
\end{equation}
where $\phi$ is the external flux \cite{Cea:2012qw} and $\lambda$ is the penetration depth. Under Ampère’s law and the definition $\mathbf{b} = \nabla \times \mathbf{a}$, one has $\mathbf{j} = (1/4\pi) \nabla \times \mathbf{b}$, which leads to an inhomogeneous differential equation for the azimuthal component of the vector potential $a_\varphi$
\begin{equation}
	\frac{d}{dx_t}{\left[\frac{1}{x_t}\frac{d}{dx_t}(x_t a_\varphi(x_t))\right]}-\frac{f^2(x_t)}{\lambda^2}a_\varphi(x_t)=-\frac{\phi f^2(x_t)}{2\pi\lambda^2x_t}.
	\label{eq:G-L-PDE}
\end{equation}

\begin{figure}[htpb]
	\begin{center}
		\includegraphics[width=0.49\textwidth]{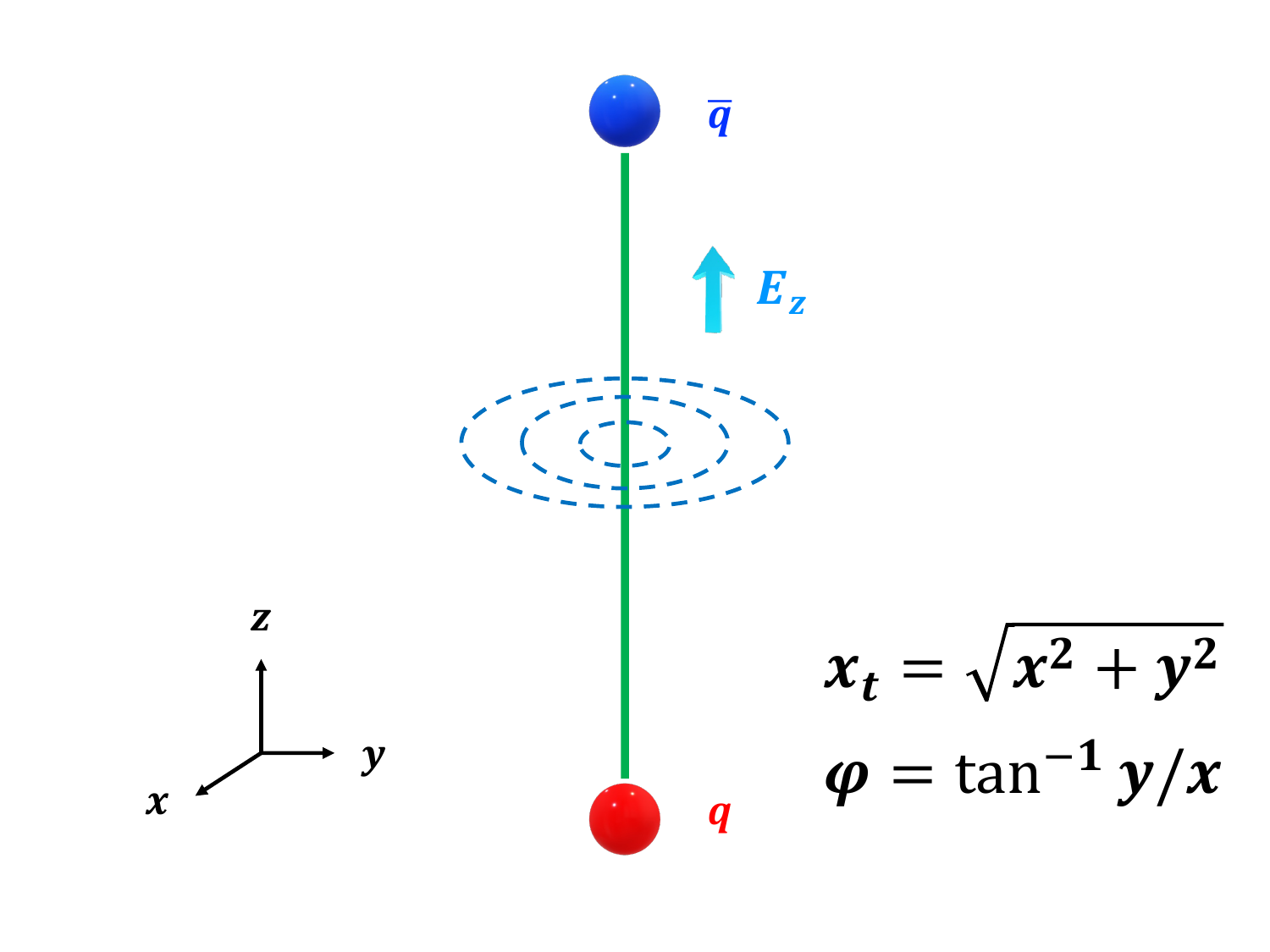}
		\caption{A schematic diagram of the coordinate system used to describe the chromoelectric field distribution between a quark and an antiquark. }
		\label{fig:qqbar}
	\end{center}
\end{figure}

An analytical solution to Eq.  (\ref{eq:G-L-PDE}) exists as
\begin{equation}
	a_\varphi=\frac{\phi}{2\pi x_t}{\left[1-\frac{RK_1(R/\lambda)}{\xi_vK_1(\xi_v/\lambda)}\right]},
	\label{eq:a_phi}
\end{equation}
 involving the modified Bessel function $K_n(x)$, yielding the corresponding expressions for the magnetic flux density and the supercurrent density as follows:
\begin{equation}
	\begin{aligned}
		&b_{z}=(\phi/2\pi\lambda\xi_{v})K_{0}(R/\lambda)/K_{1}(\xi_{v}/\lambda),\\
		&j_{\varphi}=(\phi/8\pi^{2}\lambda^{2}\xi_{v})(x_t/R)K_{1}(R/\lambda)/K_{1}(\xi_{v}/\lambda).
		\end{aligned}
	\label{eq:flux-density}
\end{equation}
Although the above formulation represents a simplified version of a superconducting model, it successfully captures the profile of the color field distribution between static quark–antiquark pairs as obtained in lattice QCD simulations \cite{Cea:2012qw,Cea:2013oba,Cea:2014hma,Cea:2014uja,Baker:2019gsi,Baker:2024peg,Baker:2024rjq}. This is primarily attributed to the dual superconductor picture, in which the flux-tube structure connecting a quark and an antiquark naturally embodies the confinement mechanism. Within this picture, the magnetic flux density in a superconductor can be interpreted as analogous to the chromoelectric field strength in QCD, both depending on the radial coordinate, and one may identify $E_z(x_t) = b_z(x_t)$. Figure \ref{fig:qqbar} presents a schematic diagram of the coordinate system used to describe the chromoelectric field distribution between a quark and an antiquark. Here, the field strength $E_z$ depends solely on the transverse coordinate, namely the radial coordinate $x_t$ in cylindrical coordinates, with its direction oriented from the quark to the antiquark.

This approach has been employed in numerous lattice studies, from which the effective parameters of the superconducting equations have been extracted by fitting to the simulated color field profiles. It is worth noting that the color field distribution between quark pairs in lattice simulations is typically obtained using the vacuum expectation value of a combination of the plaquette, Wilson loop, and Schwinger line operators, which in the continuum limit yields the field strength tensor $F_{\mu\nu}$ corresponding to the chromoelectric field.

We now return to Eq. (\ref{eq:G-L-PDE}), which is a differential equation governing the vector potential. Once $a_\varphi$ is determined, the magnetic flux density—and thus the corresponding chromoelectric field strength $E_z$—can be obtained via standard operations. Moreover, by imposing the conditions that the vector potential $a_\varphi$ vanishes both at the vortex core and at spatial infinity, Eq. (\ref{eq:G-L-PDE}) can be reformulated as an integral-differential equation (IDE), which takes the following explicit form:
\begin{equation}
	\frac{d}{d x_t}E_{z}(x_t)-\frac{f(x_t)^{2}}{\lambda^{2}}\cdot\frac{1}{x_t}\int_{0}^{x_t}sE_{z}(s)ds=-\frac{\phi f(x_t)^{2}}{2\pi\lambda^{2}x_t},
	\label{eq:G-L-IDE}
\end{equation}
with $E_z(x_t)=\frac{1}{x_t}\frac{d}{dx_t}(x_t a_\varphi(x_t))$. Both Eq. (\ref{eq:G-L-PDE}) and Eq. (\ref{eq:G-L-IDE}) can be used, either directly or indirectly, to obtain the chromoelectric field distribution and to fit the corresponding model parameters based on lattice QCD simulations. In previous work, we explored non-physics-informed neural network architectures \cite{Kou:2024hzd}, such as MLP and KAN, to analyze various chromoelectric field distributions. In particular, the KAN framework enabled us to obtain analytical expressions for the color field profiles. In this study, we adopt the framework of PINNs to analyze these two equations. By treating the color field or flux tube profiles, generated at various quark–antiquark separations \cite{Baker:2019gsi}, as input data, we apply an inverse problem approach to solve the associated PDE/ODE/IDE without relying on explicit analytical expressions, thereby extracting the effective parameters of the superconducting model \cite{clem1975simple}.

\section{PINN framework and PDE/IDE}
\label{sec:PINN}
PINNs integrate physical laws, typically expressed as PDEs, directly into the training process of neural networks. By incorporating these governing equations into the loss function, PINNs ensure that the network’s predictions not only fit the data but also adhere to the underlying physical principles. This approach enhances generalization, especially in scenarios with limited or noisy data, and enables the solution of both forward and inverse problems. PINNs serve as mesh-free alternatives to traditional numerical methods, leveraging automatic differentiation for efficient computation of derivatives, and have been successfully applied across various domains, including fluid dynamics, structural mechanics, and biomedical engineering \cite{thuerey2025physicsbaseddeeplearning}. As a neural network framework for data-driven solutions to physical problems, PINNs can be employed to solve forward problems involving PDEs with known parameters, as well as to infer parameter sets from data, which falls within the domain of inverse problems. When a PDE lacks explicit boundary condition information, the model must be trained in a manner that is driven solely by data and the governing equations (physics). Thus, training PINNs simultaneously constitutes the process of solving the associated PDEs.

Within the PINNs framework, the architectures for forward and inverse problems differ slightly, but both generally consist of an input layer, output layer, and multiple hidden layers. The interconnections among layers follow the standard structure of neural networks. Activation functions enable the network to capture nonlinear relationships, while the weights and biases are iteratively updated during training. A fully connected feedforward neural network is constructed from linear summations combined with activation functions. For networks employing differentiable activation functions, derivatives of the output with respect to the input can be computed via backward chain. This technique is known as AD \cite{Baydin:2015tfa}, it is a built-in function that can be directly called in the popular ML backends such as Pytorch \cite{paszke2017automatic,paszke2019pytorch}, Tensorflow \cite{abadi2016tensorflow}, Jax \cite{lin2024automaticfunctionaldifferentiationjax} and Paddle \cite{paddlepaddle2025}. 

\begin{figure}[htpb]
	\begin{center}
		\includegraphics[width=0.49\textwidth]{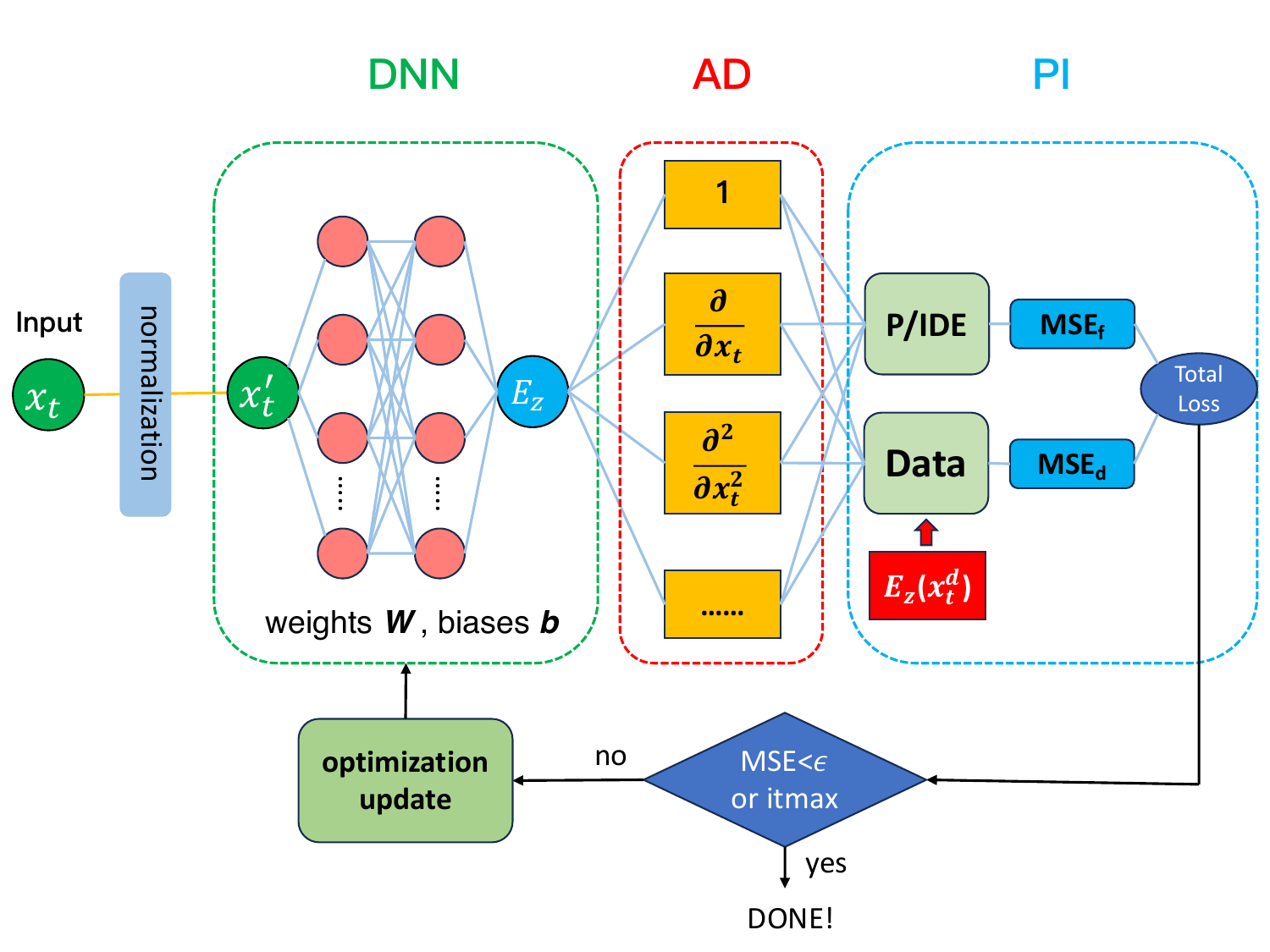}
		\caption{The framework of PINNs to solve the inverse problem of PDE/ODE/IDE.}
		\label{fig:pinn}
	\end{center}
\end{figure}

As an example, we consider the PINNs framework for solving inverse problems, and examine the implementation of Eq.~(\ref{eq:G-L-PDE}) within this context. The ODE can be expressed as follows:
\begin{equation}
	\mathcal{F}[\omega_m; x_t, \partial_{x_t}, a_\varphi]\equiv0,\quad x_t\in[0,\infty],
\end{equation}
where $\omega_m$ denote the all parameters of ODE such as $\lambda,\xi_v,\phi$, and operator $\mathcal{F}$ represents the residual obtained by subtracting the right-hand side of Eq.~(\ref{eq:G-L-PDE}) from its left-hand side. The PINNs framework for solving the inverse problem of PDE is illustrated in Fig. \ref{fig:pinn}, now the all unknow parameters $\omega_m$ as hyper-parameters are trained together with the DNNs. We emphasize that the data used in this study are obtained from lattice simulations of $E_z(x_t)$; however, this quantity can also be derived from the vector potential $a_\varphi(x_t)$. Therefore, we consider the both to be in one-to-one correspondence.

The input $x_t$ of the PINNs corresponds to the coordinates of the training points. Since we are addressing an inverse problem, initial and boundary condition sampling points typically available in forward problems are absent; only sampling points within the domain of the equation are considered. A fully connected feedforward DNN can be used to compute the predicted solution of the equation at the coordinates $x_{t,i}^d$, denoted as $E_z^{{pred}}(x_{t,i}^d;\omega_m,\theta)$. This prediction is then compared to the measured data $E_z^d(x_{t,i}^d)$, where $\theta$ represents the hyperparameters of the DNN, and $\omega_m$ denotes the unknown parameters in the PDE/ODE/IDE that are to be inferred. In the analysis of inverse problems, typically only the governing equations and measured data are available. In such cases, the physical information is composed of two components: the residuals of the PDE/ODE/IDE and the residuals corresponding to the measured data. In the absence of boundary and initial conditions as physical constraints, the total loss function $MSE_{\text{total}}$ can be expressed as \cite{raissi2019physics}
\begin{equation}
	\begin{aligned}
		&MSE_{total}=MSE_d+MSE_f,\\
		&MSE_d=\frac{1}{N_d}\sum_{i=1}^{N_d}\left|E_z^{pred}\left(x_{t,i}^d;\theta,\omega_m\right)-E_z^d\left(x_{t,i}^d\right)\right|^2,\\
		&MSE_f=\frac{1}{N_f}\sum_{i=1}^{N_f}\left|\mathcal{F}[\omega_m; x_{t,i}^f, \partial_{x_{t,i}^f}, E_z]\right|^2,
	\end{aligned}
	\label{eq:totalMSE}
\end{equation}
where $MSE_d$ and $MSE_f$ are the mean square error of residuals of the measurement data and the PDE/ODE/IDEs, respectively. $N_d$ is the number of the training points and $N_f$ is the number of collocation points randomly sampled in the equation domain.

\begin{figure}[htpb]
	\begin{center}
		\includegraphics[width=0.49\textwidth]{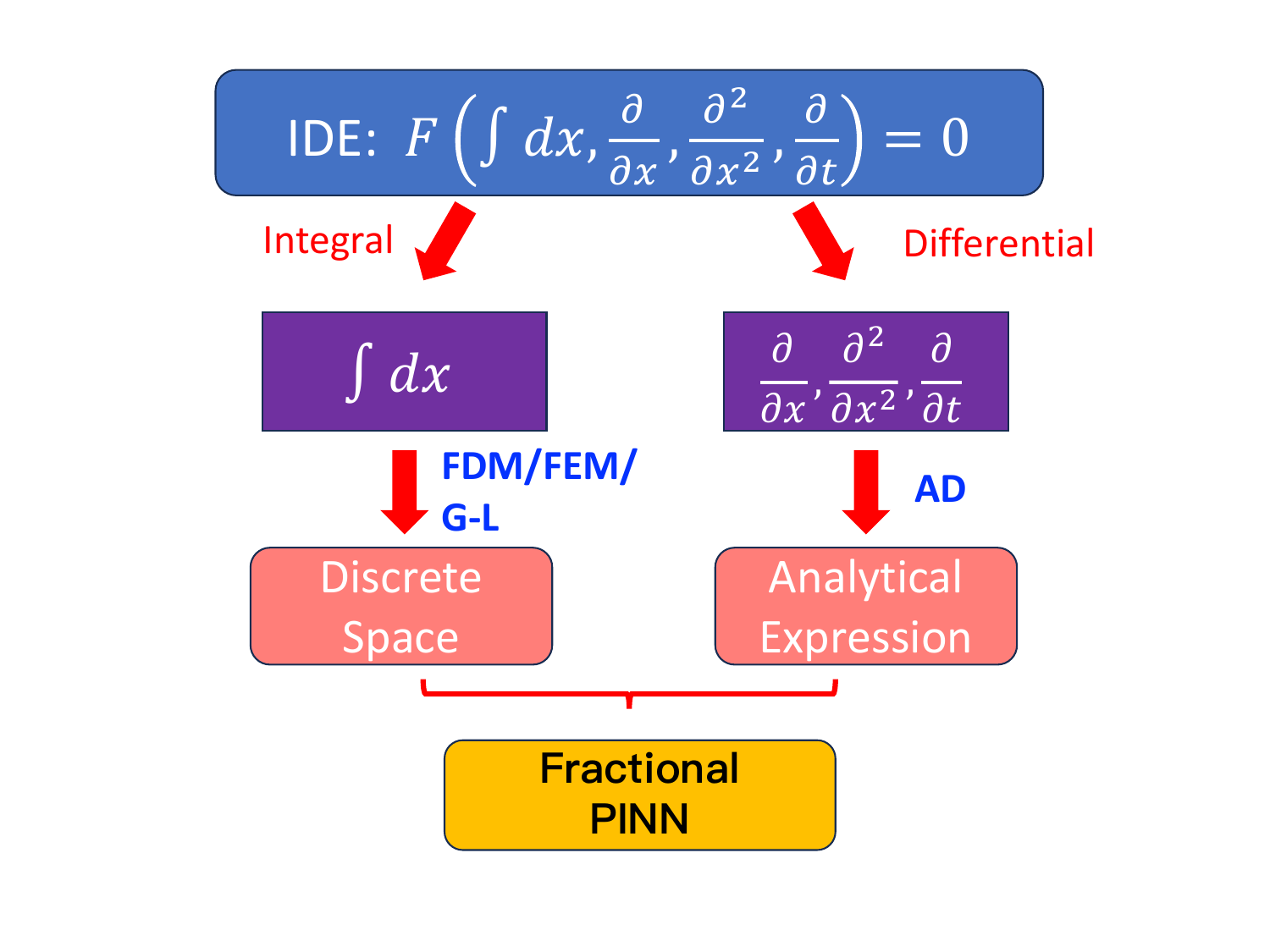}
		\caption{A schematic diagram of the discretization strategy for solving integro-differential equations within the PINNs framework is shown \cite{Pang_2019}. While the differential terms can be handled via AD, the integral components require appropriate discretization schemes, such as the finite element method (FEM), the finite difference method (FDM), or Gauss–Legendre quadrature (G-L).}
		\label{fig:fpinn}
	\end{center}
\end{figure}

We now turn to the case of handling IDEs within the PINNs framework. In this scenario, AD remains applicable for the differential operators in the equations; however, the integral operators must be incorporated through numerical techniques, specifically by discretizing the integrals (see Fig. \ref{fig:fpinn}) \footnote{Here,``fractional" refers to the embedding of discretized integrals into the PINN, and does not imply that our work employs fractional-order differential operators. The term``fractional" is retained only as an original label \cite{Pang_2019}.}. This transformation effectively recasts the IDEs into ODEs. Prior studies addressing IDEs using discretization schemes within PINNs, such as those in Refs. \cite{Pang_2019,yuan2022pinn}, are highly instructive—particularly the auxiliary PINNs strategy proposed in Ref. \cite{yuan2022pinn}, which reformulates complex IDEs into systems of PDEs, thereby expanding the flexibility in designing loss functions. In the present work, we adopt a discretization scheme based on Gaussian quadrature to treat Eq.~(\ref{eq:G-L-IDE}), rewriting the integral in a discrete form as follows
\begin{equation}
	\int_0^x y(s)ds\simeq \sum_{i=1}^{n}\alpha_iy(s_i(x)),
\end{equation}
where $s_i(x)$ and $\alpha_i$ are the known integral nodes and weighting factors, respectively. In this work, we set the 20th-order Gauss Legendre quadreture for the IDE (\ref{eq:G-L-IDE}).

\section{results and discusstion}
\label{sec:results}
The core objective of this work is to use the color field distributions $E_z(x_t)$, obtained from lattice simulations at various color source separations $d$, as training data to solve the equations arising from the dual superconductivity picture—specifically, the ODE~(\ref{eq:G-L-PDE}) and the IDE~(\ref{eq:G-L-IDE}). We assume, under certain conditions, that these equations are physically compatible. Each color field distribution corresponding to a given $d$ is trained independently, with 20$\%$ of the primary sampling points reserved as a test set. 

For comparison, in addition to the lattice simulation data, we also consider the Clem parametrization from Ref. \cite{Baker:2019gsi} as a benchmark for the inverse problem of recovering equation parameters. The parametrization employed in Ref. \cite{Cea:2012qw} is a modified version of Eq.~(\ref{eq:flux-density}), where $b_z = E_z$, namely:
\begin{equation}
	\begin{aligned}
		E_z(x_t)&=\frac{\phi}{2\pi}\frac{1}{\lambda\xi_v}\frac{K_0(R/\lambda)}{K_1(\xi_v/\lambda)}\\
		&=\frac{\phi}{2\pi}\frac{\mu^2}{\alpha}\frac{K_0[(\mu^2x_t^2+\alpha^2)^{1/2}]}{K_1[\alpha]}
	\end{aligned}
	\label{eq:para}
\end{equation}
with
\begin{equation}
	\mu=\frac{1}{\lambda},\quad\frac{1}{\alpha}=\frac{\lambda}{\xi_v}.
	\label{eq:para-transf}
\end{equation}
The second line of Eq. (\ref{eq:para}) offers a notable advantage: it allows for fitting procedures that include the transverse coordinate $x_t = 0$. Furthermore, the parameters $\mu$ and $\alpha$ represent the penetration depth and the ratio of the variational core radius to the penetration depth, respectively. The Ginzburg–Landau parameter $\kappa$, which characterizes the type of superconductivity, is defined by the following relation \cite{clem1975simple}
\begin{equation}
	\kappa=\frac{\sqrt{2}}{\alpha}[1-K_0^2(\alpha)/K_1^2(\alpha)]^{1/2},
\end{equation}
and can be used in conjunction with the relation $\kappa = \lambda / \xi$ to determine the coherence length $\xi$.

We further present the string tension $\sqrt{\sigma_T}$ between the quark and antiquark, as well as the average width $\sqrt{w^2}$ of the resulting color flux tube, both computed from the solutions of the trained equations or the inferred color field distributions. These quantities are defined as follows
\begin{equation}
	\sqrt{\sigma_T}=\sqrt{\int d^2x_t\frac{E_z^2(x_t)}{2}},\quad \sqrt{w^2}=\sqrt{\frac{\int d^2x_tx_t^2E_z(x_t)}{\int d^2x_tE_z(x_t)}}.
	\label{eq:st-width}
\end{equation}
\subsection{ODE case}
\label{subsec:ODE}
\begin{figure*}[htpb]
	\begin{center}
		\includegraphics[width=0.98\textwidth]{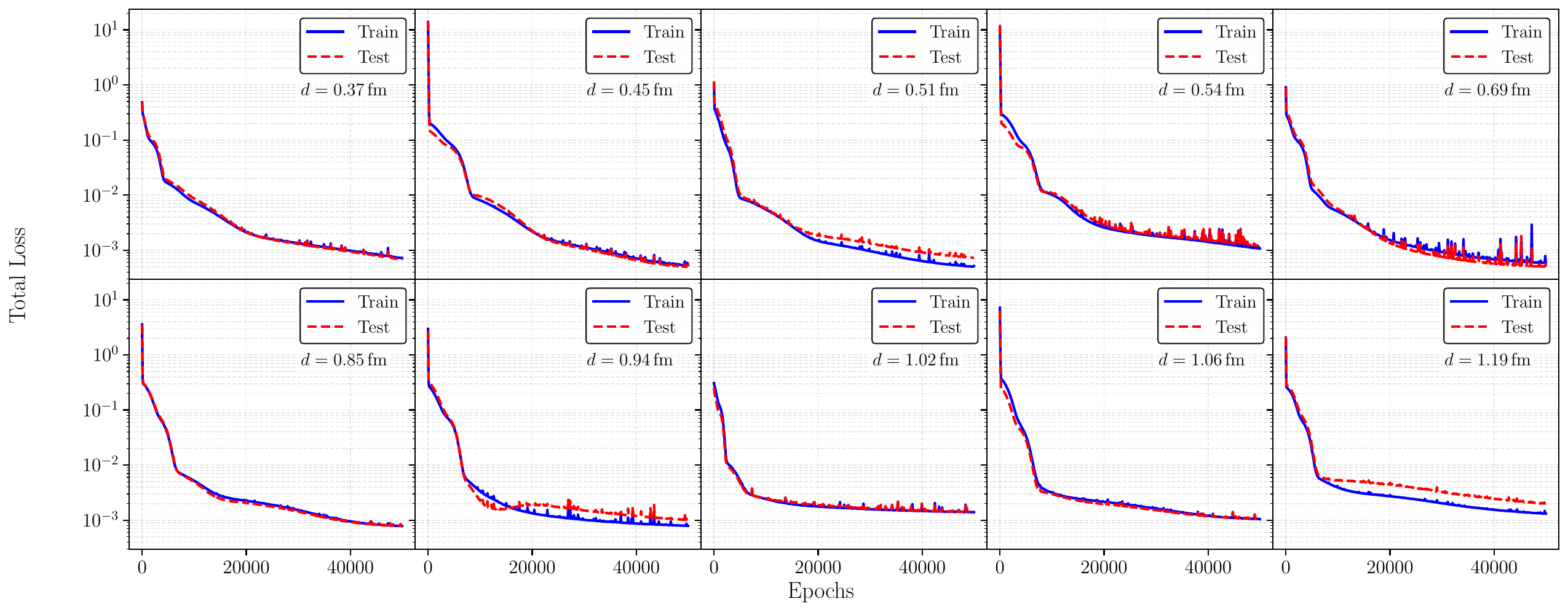}
		\caption{The loss function curves corresponding to the solution of the ODE using the PINNs framework are presented, with the testing set (red) constructed by randomly selecting 20$\%$ of the samples from the training dataset (blue).}
		\label{fig:ODE-loss}
	\end{center}
\end{figure*}

We first present the results obtained by solving the original ODE (\ref{eq:G-L-PDE}). Since the lattice simulations provide the chromoelectric field $E_z$, whereas the solution of the ODE corresponds to the gauge potential $a_\varphi$, the data residual must account for an additional AD step, given by $E_z(x_t) = \frac{1}{x_t} \frac{d}{dx_t} (x_t a_\varphi(x_t))$. The neural network architecture is configured as [1,70,70,70,1], meaning that each of the three hidden layers contains 70 neurons. A learning rate of $10^{-5}$ is employed, and the maximum number of training iterations is set to 50,000. In addition to the hyperparameters of the neural network itself, the unknown parameters of the equation, such as $\phi$, $\lambda$, and $\xi_v$, are also included in the optimization process. The lattice simulation provides ten sets of $E_z(x_t)$, each corresponding to the distribution of the chromoelectric field at quark-antiquark source separations of $d = 0.37$, $0.45$, $0.51$, $0.54$, $0.69$, $0.85$, $0.94$, $1.02$, $1.06$, and $1.19$ fm. Each set of $E_z(x_t)$ corresponding to a given $d$ is trained individually, and the total loss functions for both the training (blue) and testing (red) datasets are presented in Fig. \ref{fig:ODE-loss}. It is important to emphasize that we define the total loss $MSE_{total}$ as a weighted sum, where the ratio between the ODE loss and the data loss is set to $0.2:1$. This choice reflects the fact that the data are not continuous and exhibit certain non-physical fluctuations; hence, the data loss must retain a non-zero weight. According to the training results, both the ODE loss and the data loss, for both the training and test sets, stabilize at the level of $10^{-4} \sim 10^{-5}$ after 50,000 iterations.

\begin{figure*}[htpb]
	\begin{center}
		\includegraphics[width=0.98\textwidth]{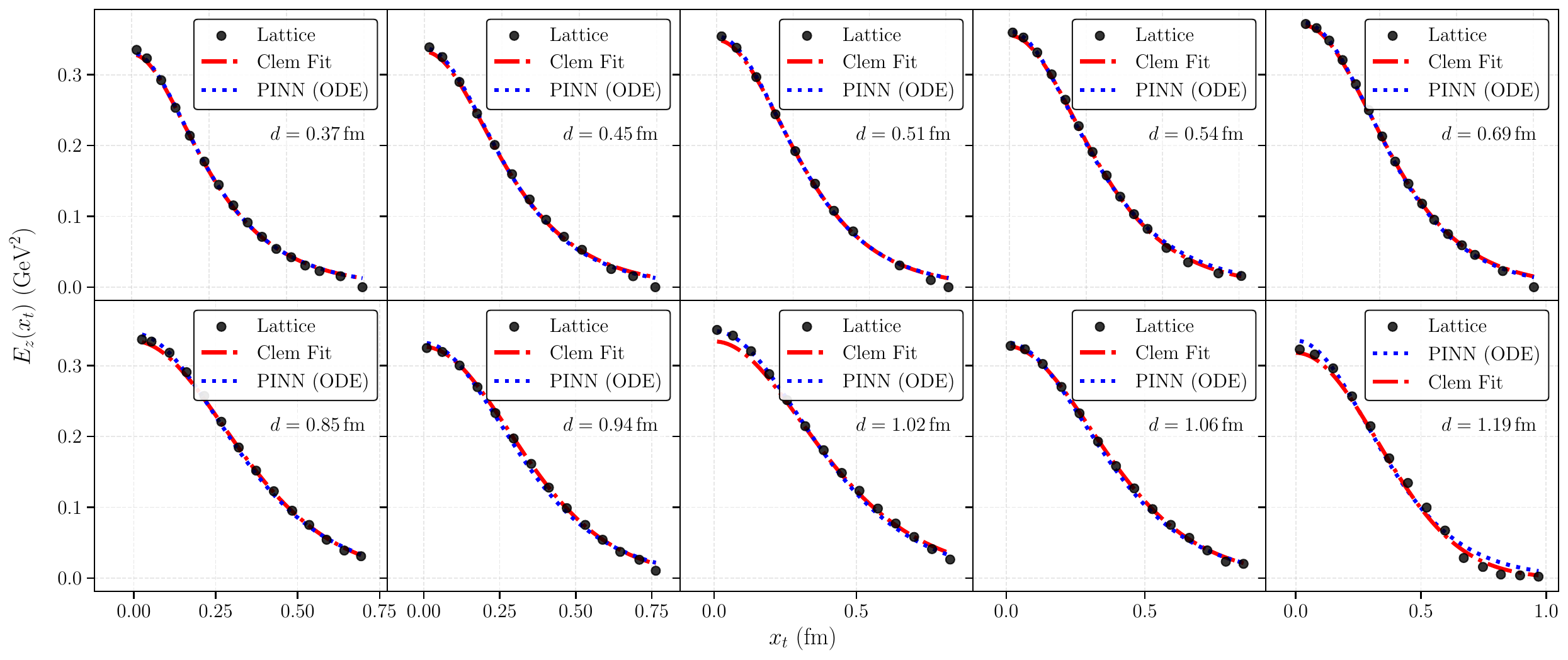}
		\caption{The chromoelectric field distributions corresponding to the solution of the ODE using the PINNs framework are presented. The PINNs results (blue curves) are compared with Clem fit (red curves) and lattice simulations (black solid sircles) \cite{Baker:2019gsi}.}
		\label{fig:field-PINN-ODE}
	\end{center}
\end{figure*}

A comparison between the chromoelectric field distributions predicted by the PINNs framework for solving the ODE, the lattice simulation results, and the Clem ansatz is presented in Fig. \ref{fig:field-PINN-ODE}. It can be observed that the blue curve, representing the PINNs prediction, and the red curve, corresponding to the Clem parameterization, are both in good agreement with the black data points from the lattice simulation.

It is known that the discrete field distributions obtained from lattice simulations can be effectively fitted using the exact solution of the ODE expressed through the Clem parameterization (\ref{eq:para}). The referenced work \cite{Baker:2019gsi} also provides the field distribution parameters corresponding to different values of $d$. By employing the PINNs framework to address the inverse problem, the equation parameters can be treated as unknowns embedded within the feedforward model. Consequently, while training to solve the ODE, we are able to infer the equation parameters, specifically $\xi_v$, $\phi$, and $\lambda$ in Eq. (\ref{eq:G-L-PDE}). For comparison with the lattice-based analysis, we also performed a parameter transformation to extract the inferred values of $\phi$, $\mu$, and $\alpha$ in Eqs. (\ref{eq:para}) and (\ref{eq:para-transf}). 

Table \ref{table:para-ODE} presents the equation parameters extracted by fitting the lattice simulation results using the Clem parameterization, alongside those inferred by solving the ODE with the PINNs framework. The method adopted in this work does not rely on an explicit analytical expression for the solution; instead, the optimal parameters are entirely determined by the neural network.
\begin{table}[htbp]
	\caption{\label{table:para-ODE}Comparison between lattice fitting \cite{Baker:2019gsi} and PINNs (ODE) inference for parameters $\phi$, $\mu$, and $\alpha$ at various values of $d$.}
	\begin{ruledtabular}
		\begin{tabular}{l
				D{.}{.}{3.3} D{.}{.}{3.3} D{.}{.}{3.3}
				D{.}{.}{3.3} D{.}{.}{3.3} D{.}{.}{3.3}}
			\multicolumn{1}{c}{$d$ (fm)} &
			\multicolumn{3}{c}{Lattice fitting} &
			\multicolumn{3}{c}{PINNs inference} \\
			\cmidrule(lr){2-4} \cmidrule(lr){5-7}
			& \multicolumn{1}{c}{$\phi$} & \multicolumn{1}{c}{$\mu$ (fm$^{-1}$)} & \multicolumn{1}{c}{$\alpha$} &
			\multicolumn{1}{c}{$\phi$} & \multicolumn{1}{c}{$\mu$ (fm$^{-1}$)} & \multicolumn{1}{c}{$\alpha$} \\
			\midrule
			0.370 & 3.474 & 4.999 & 1.192 & 3.489 & 4.966 & 1.168 \\
			0.450 & 3.830 & 5.300 & 1.550 & 3.746 & 5.546 & 1.659 \\
			0.510 & 4.028 & 6.039 & 2.141 & 4.068 & 5.760 & 1.898 \\
			0.540 & 4.370 & 5.710 & 2.020 & 4.650 & 4.751 & 1.352 \\
			0.690 & 4.500 & 6.250 & 2.470 & 4.388 & 6.470 & 2.570 \\
			0.850 & 5.400 & 6.700 & 4.000 & 5.511 & 5.429 & 2.441 \\
			0.940 & 5.200 & 7.800 & 5.500 & 5.215 & 5.601 & 2.579 \\
			1.020 & 8.000 & 4.400 & 2.400 & 7.699 & 4.587 & 2.407 \\
			1.060 & 6.600 & 6.000 & 4.000 & 6.520 & 5.034 & 2.602 \\
			1.190 & 5.500 & 81.000 & 700.000 & 6.151 & 5.714 & 3.228 \\
		\end{tabular}
	\end{ruledtabular}
\end{table}

\subsection{IDE case}
\label{subsec:IDE}
For the inverse problem of solving the IDE in a data-driven framework, where lattice simulation data \cite{Baker:2019gsi} are used as input to Eq. (\ref{eq:G-L-IDE}), we adopt a DNN architecture with the structure $[1,64,64,64,1]$, in which each of the three hidden layers contains 64 neurons. The learning rate is set to $5\times 10^{-4}$. In addition to the intrinsic hyperparameters of the neural network, the unknown parameters appearing in the equation, such as $\phi$, $\lambda$, and $\xi_v$, are also included in the training and optimization process. The maximum number of training iterations is set to $2\times 10^4$, and the loss function is monitored as a function of the iteration number. 

\begin{figure*}[htpb]
	\begin{center}
		\includegraphics[width=0.98\textwidth]{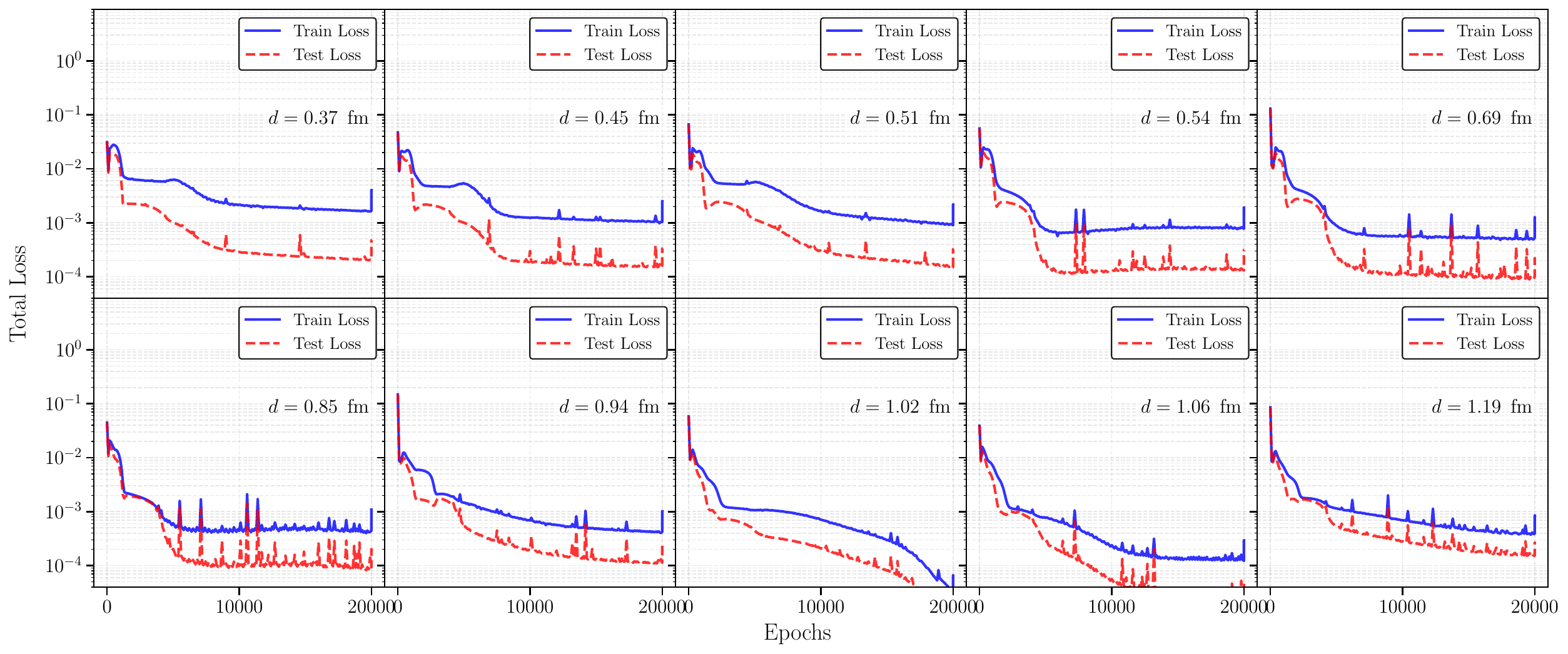}
		\caption{The loss function curves corresponding to the solution of the IDE using the PINNs framework are presented, with the testing set (red) constructed by randomly selecting 20$\%$ of the samples from the training dataset (blue).}
		\label{fig:IDE-loss}
	\end{center}
\end{figure*}

As mentioned earlier, the lattice simulation provides ten sets of $E_z(x_t)$ data, each corresponding to a specific distance between the quark and antiquark color sources, with $d =$ 0.37, 0.45, 0.51, 0.54, 0.69, 0.85, 0.94, 1.02, 1.06, 1.19 fm. Each color field distribution corresponding to a given value of $d$ is trained independently. The total loss functions for both the training (blue) and testing (red) sets are presented in Fig. \ref{fig:IDE-loss}. It is important to emphasize that we define the total loss $MSE_{total}$ as a weighted sum, where the ratio between the IDE loss and the data loss is set to $0.4:1$. This choice reflects the fact that the data are not continuous and exhibit certain non-physical fluctuations; hence, the data loss must retain a non-zero weight. According to the training results, both the IDE loss and the data loss, for both the training and test sets, stabilize at the level of $10^{-4} \sim 10^{-5}$ after 20,000 iterations.

\begin{figure*}[htpb]
	\begin{center}
		\includegraphics[width=0.98\textwidth]{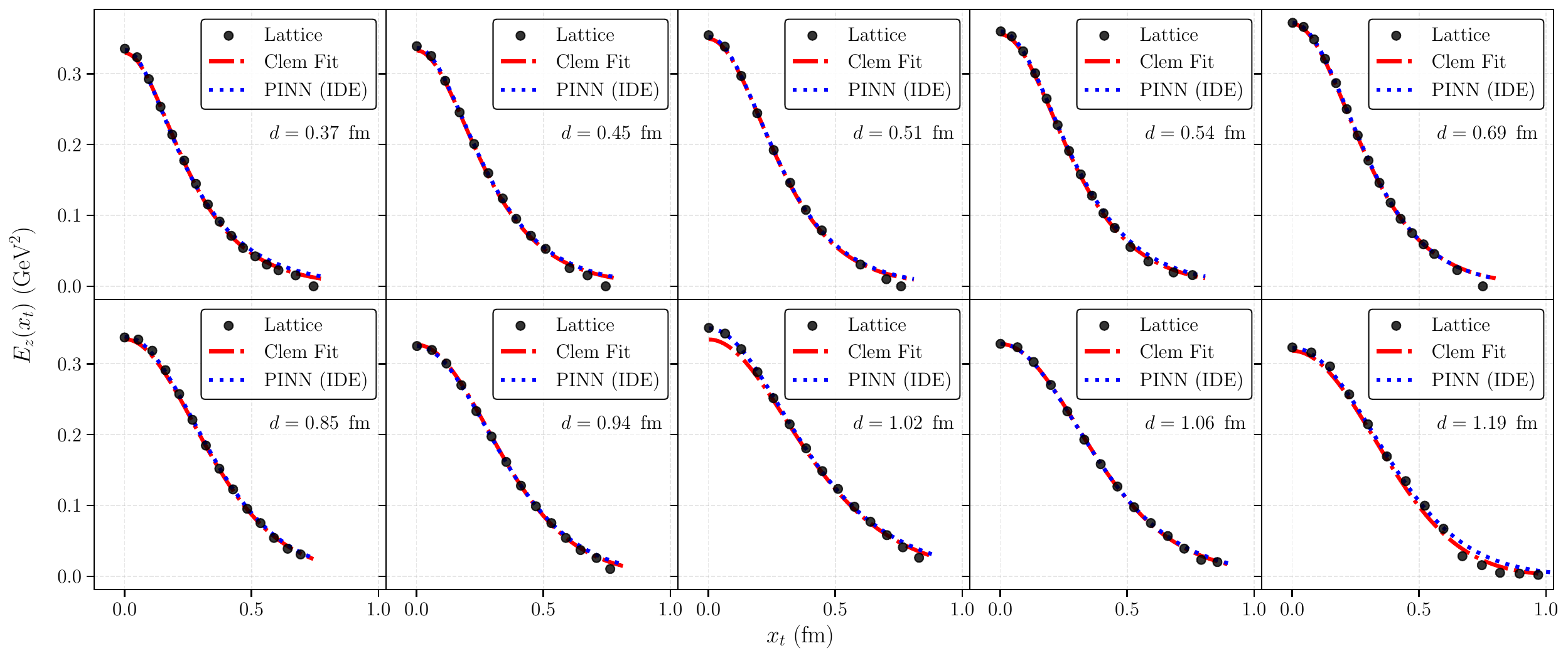}
		\caption{The chromoelectric field distributions corresponding to the solution of the IDE using the PINNs framework are presented. The PINNs results (blue curves) are compared with Clem fit (red curves) and lattice simulations (black solid sircles) \cite{Baker:2019gsi}.}
		\label{fig:field-PINN-IDE}
	\end{center}
\end{figure*}

The predicted color field distributions $E_z(x_t)$ obtained from training at various source separations $d$ are shown in Fig. \ref{fig:field-PINN-IDE}, with lattice simulation results \cite{Baker:2019gsi} included for comparison. Additionally, we compare these with the Clem parametrization (\ref{eq:para}) fitted to the lattice data. The blue curves represent the solutions of the IDE obtained through our training framework, the red curves correspond to the Clem parametrized results, and the black data points are from the lattice simulations. It is evident that the color field profiles derived from solving the superconducting equations using the PINNs framework are in good agreement with both the lattice data and the fitted parameterization, all capturing the expected decay of the color field as a function of the transverse spatial coordinate $x_t$.

Similarly, we have also inferred the relevant parameters in the process of solving the IDE. The approach to parameter inference follows the same procedure as for the ODE, where the parameters $\xi_v, \phi, \lambda$ are embedded into the feedforward model and inferred during the equation-solving process. For comparison with the lattice-based results, we performed a parameter transformation to obtain the corresponding inferred values of $\phi, \mu, \alpha$. It is important to note that the superconducting equation in the IDE form requires discretization of the integral term, which is then incorporated into the residual of the equation constraint. Table \ref{table:para-IDE} presents a comparison between the parameters inferred by solving the IDE inverse problem using PINNs and those obtained from the lattice-based analysis.
\begin{table}[htbp]
	\caption{\label{table:para-IDE}Comparison between lattice fitting \cite{Baker:2019gsi} and PINNs (IDE) inference for parameters $\phi$, $\mu$, and $\alpha$ at various values of $d$.}
	\begin{ruledtabular}
		\begin{tabular}{l
				D{.}{.}{3.3} D{.}{.}{3.3} D{.}{.}{3.3}
				D{.}{.}{3.3} D{.}{.}{3.3} D{.}{.}{3.3}}
			\multicolumn{1}{c}{$d$ (fm)} &
			\multicolumn{3}{c}{Lattice fitting} &
			\multicolumn{3}{c}{PINNs inference} \\
			\cmidrule(lr){2-4} \cmidrule(lr){5-7}
			& \multicolumn{1}{c}{$\phi$} & \multicolumn{1}{c}{$\mu$ (fm$^{-1}$)} & \multicolumn{1}{c}{$\alpha$} &
			\multicolumn{1}{c}{$\phi$} & \multicolumn{1}{c}{$\mu$ (fm$^{-1}$)} & \multicolumn{1}{c}{$\alpha$} \\
			\midrule
			0.370 & 3.474 & 4.999 & 1.192 & 3.663 & 4.531 & 0.966 \\
			0.450 & 3.830 & 5.300 & 1.550 & 3.991 & 4.970 & 1.368 \\
			0.510 & 4.028 & 6.039 & 2.141 & 4.240 & 5.341 & 1.671 \\
			0.540 & 4.370 & 5.710 & 2.020 & 4.474 & 5.541 & 1.910 \\
			0.690 & 4.500 & 6.250 & 2.470 & 4.616 & 6.084 & 2.371 \\
			0.850 & 5.400 & 6.700 & 4.000 & 5.632 & 6.154 & 3.419 \\
			0.940 & 5.200 & 7.800 & 5.500 & 5.470 & 6.172 & 3.487 \\
			1.020 & 8.000 & 4.400 & 2.400 & 8.458 & 3.799 & 1.702 \\
			1.060 & 6.600 & 6.000 & 4.000 & 6.718 & 5.546 & 3.409 \\
			1.190 & 5.500 & 81.000 & 700.000 & 6.200 & 8.679 & 8.411 \\
		\end{tabular}
	\end{ruledtabular}
\end{table}

\subsection{Discusstions}
\begin{figure*}[htpb]
	\begin{center}
		\includegraphics[width=0.49\textwidth]{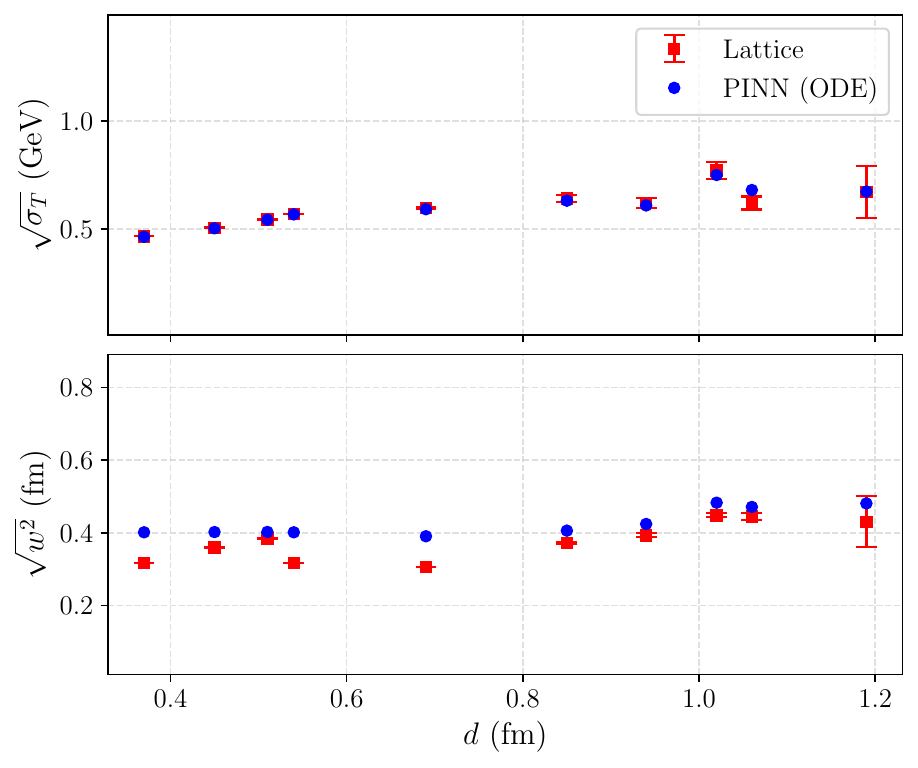}
		\includegraphics[width=0.49\textwidth]{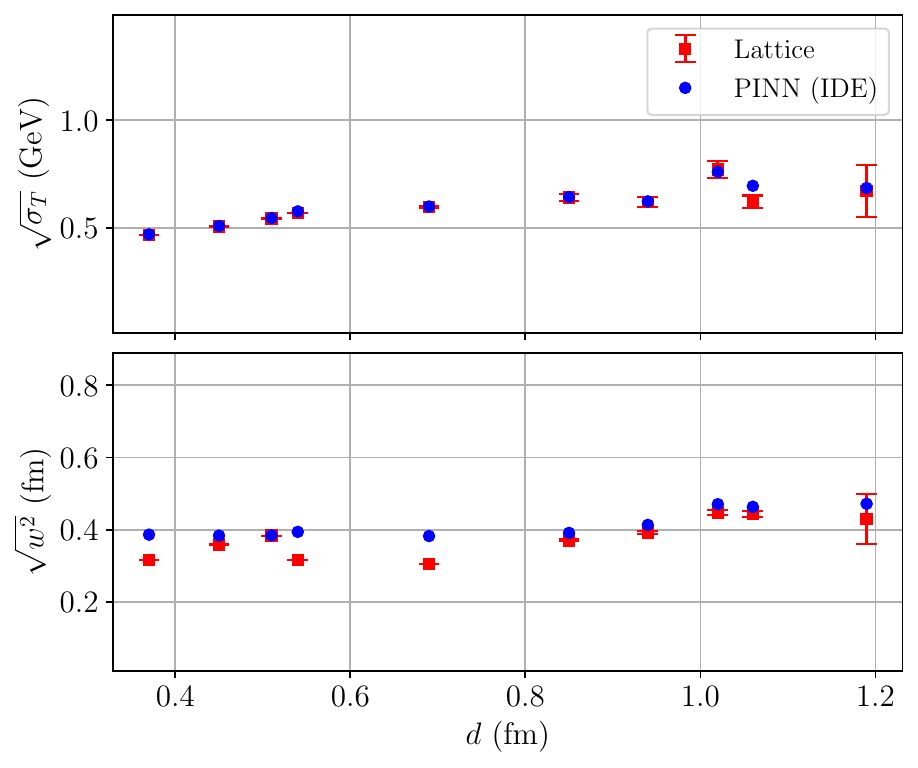}
	\end{center}
	\caption{The results for the string tension and the root-mean-square width of the flux tube are shown in the figure. The blue and red data points represent the lattice simulation results and the PINNs predictions, respectively (left: ODE-based, right: IDE-based).}
	\label{fig:st-width}
\end{figure*}

We have presented the results obtained using the PINNs framework to solve the dual superconducting equations for color field distributions in both the ODE and IDE formulations. This includes comparisons between the predicted solutions, lattice simulation results, and the Clem parametrization. Furthermore, we have demonstrated the capability of the inverse problem approach to infer the unknown parameters in the governing equations. Overall, the results show that PINNs can effectively be employed to solve the dual superconducting equations in both their ODE and IDE forms.
In addition, the extracted color field distributions $E_z(x_t)$ were used to compute physical observables associated with the flux tube structure, namely the string tension and the root-mean-square width, as defined in Eq. (\ref{eq:st-width}). These computed quantities are compared with corresponding lattice results, as shown in Fig. \ref{fig:st-width}. It can be observed that the string tension values obtained from both the ODE- and IDE-based approaches exhibit excellent agreement with the lattice data. However, slight discrepancies arise in the predictions of the root-mean-square width.
Two factors may contribute to this difference. First, the range of input data in $x_t$ affects the expressivity and generalization capability of the PINNs, and large-$x_t$ fluctuations may lead to inaccuracies in the integration of higher-order moments. Second, since the PINNs model is also trained on discretized data, it may yield less accurate results when evaluating integrals involving strongly oscillatory behavior.

\begin{figure*}[htpb]
	\begin{center}
		\includegraphics[width=0.49\textwidth]{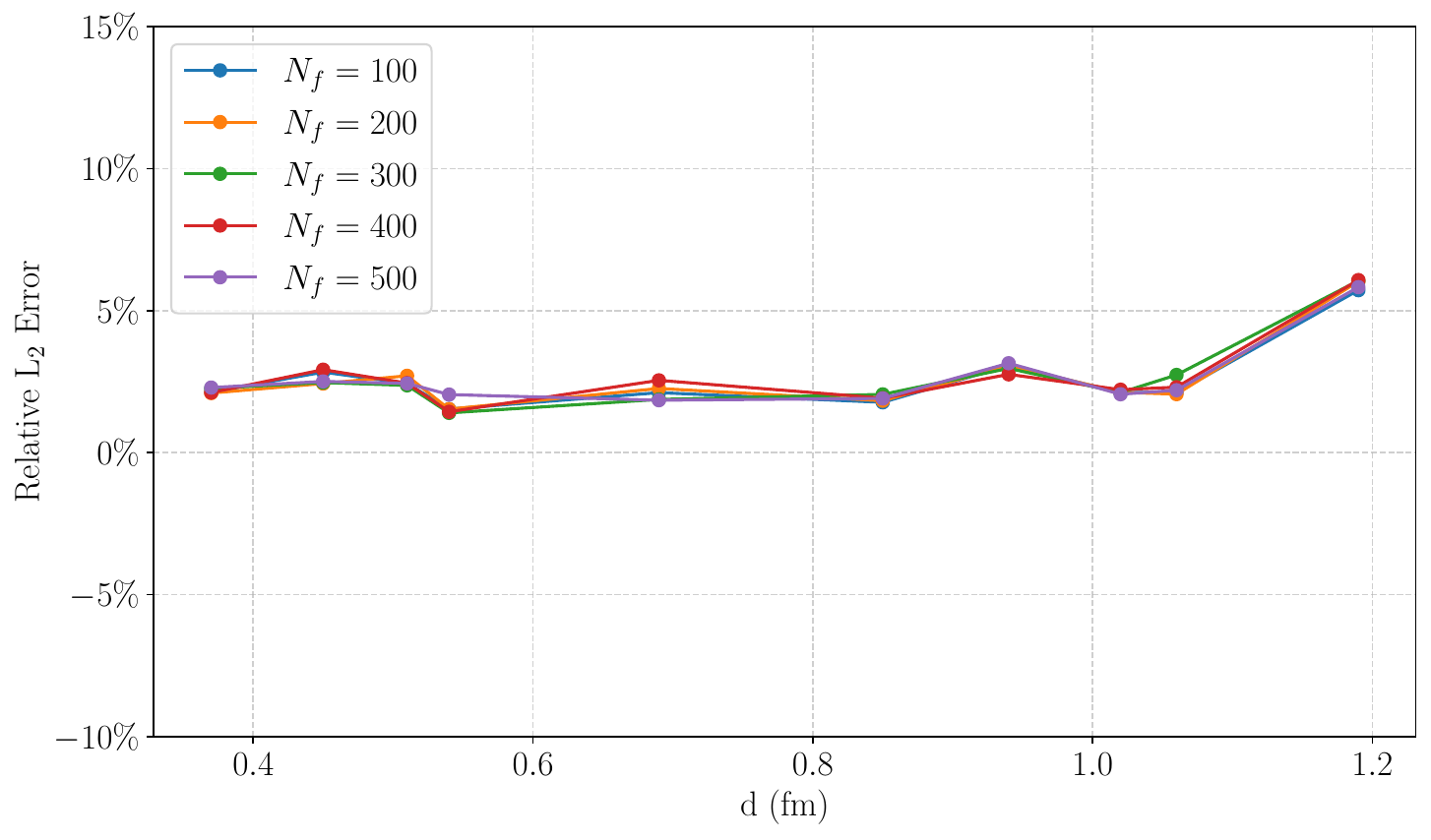}
		\includegraphics[width=0.49\textwidth]{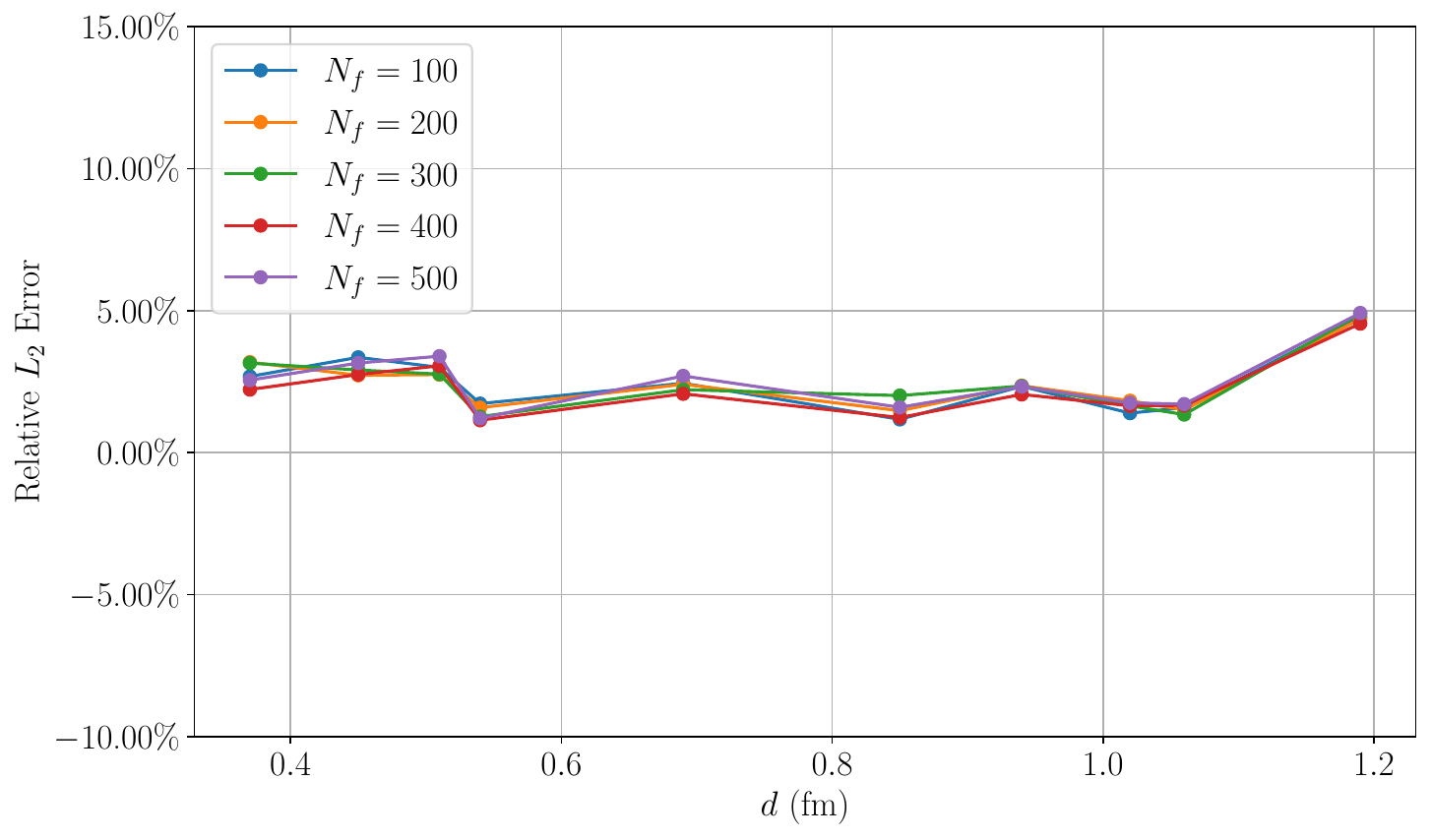}
	\end{center}
	\caption{The relative $L_2$ training errors corresponding to primary collocation point numbers $N_f = 100$, $200$, $300$, $400$, and $500$ (left: ODE-based, right: IDE-based).}
	\label{fig:L2error}
\end{figure*}

We now examine the robustness of the PINNs approach in solving the dual superconducting equations in a data-driven manner. To quantify the deviation between the PINNs-predicted color field distribution and the corresponding lattice simulation results, we employ the relative $L_2$ error, defined as
\begin{equation}
	L_2\mathrm{~error}=\frac{\sqrt{\sum_{i=1}^M\left|E_{z,pred}^i-E_{z,exact}^i\right|^2}}{\sqrt{\sum_{i=1}^M\left|E_{z,exact}^i\right|^2}}.
	\label{eq:L2error}
\end{equation}
To assess the stability of the neural network configuration, we evaluate the impact of varying the number of primary collocation points used in the equation. Specifically, we consider values of 100, 200, 300, 400, and 500 sampling points $N_f$. The resulting relative $L_2$ errors between the trained predictions and the lattice data are shown in the Fig. \ref{fig:L2error}, including results obtained using both the ODE and IDE formulations. Since the number of collocation points primarily affects the residuals in the physical PDE constraints, we treat it as the central parameter in our analysis. It is evident that, across all cases—regardless of the color field distribution corresponding to each $d$ or the choice between ODE and IDE approaches—the relative $L_2$ error remains below approximately 5$\%$.

We also explored different neural network configurations, such as increasing the number of hidden layers while keeping the number of neurons per layer fixed. The resulting relative $L_2$ errors remained nearly unchanged, indicating that the neural architecture previously adopted is effective for training the dual superconducting equation. We note that the chosen PDE/data loss weights—0.2:1 for the ODE and 0.4:1 for the IDE—were arrived at through extensive hyperparameter tuning rather than arbitrary selection. Specifically, we conducted grid-style sensitivity studies over weight values ranging from 0.1 to 0.8, retraining the PINN models in each case and evaluating performance in terms of validation loss, RMS error, and robustness of inferred parameters. This systematic search consistently revealed that the selected ratios minimize joint loss and yield the most stable parameter estimates (variations within approximately 5$\%$). Importantly, the IDE formulation, which incorporates numerical quadrature (e.g., Gauss–Legendre), alters the scale and smoothness of the PDE residuals, necessitating a larger PDE weight to maintain balanced training. Achieving the optimal balance between fitting lattice data and enforcing physical constraints is a nontrivial multi-objective optimization problem, akin to finding a ``Pareto-optimal” point across competing loss components \cite{IEEE:2023}.

Finally, we analyze the discrepancies in the inferred parameters. It is important to emphasize that the Clem fit employed in lattice studies is itself an analytical solution to the governing equation, and the fitted parameters accurately reproduce the color field distribution. In contrast, our approach does not rely on the analytical form of the solution; instead, the parameters are inferred solely from the equation structure and the lattice color field data through the residuals during training. Although the inferred parameters obtained via our method differ from those derived through Clem parameterization—both in the ODE and IDE scenarios—the resulting color field distributions still reasonably reproduce the discrete lattice data.
As the colour field profile depends on three parameters, the analysis indicates that direct fitting is not the sole avenue for constraining their values. This is particularly evident at $d = 1.19$ fm, where the fitted values of $\mu$ and $\alpha$ deviate significantly from their inferred counterparts \cite{Baker:2019gsi}, yet both parametrizations reproduce the colour field distribution in agreement with the data.

\section{conclusion}
\label{sec:conlusion}

In this study, we employ the PINNs framework to investigate the flux tube structure (colour field distribution) between static quark–antiquark pairs, guided by the dual superconductivity paradigm. Discrete lattice QCD data at varying source separations are used to solve the dual superconducting equations, formulated in both ordinary and integro-differential forms. The PINNs architecture incorporates automatic differentiation for the differential formulation and adopts a fractional PINNs strategy for the integro-differential representation of the Ginzburg–Landau equations. Our results show that accurate solutions can be achieved with minimal dependence on boundary conditions; the governing equations, embedded as residual terms in the loss function, and input field data suffice to constrain the system. Unknown physical parameters intrinsic to the equations are concurrently inferred during training via backpropagation, enabling a data-driven characterization of the dual superconducting vacuum.

The PINNs framework demonstrates notable efficacy in addressing inverse problems, particularly in scenarios where explicit initial or boundary conditions are inaccessible. In this study, the PINNs-reconstructed colour field distributions accurately recover the solutions of both the differential and integro-differential formulations, underscoring a strong correspondence between non-Abelian gauge dynamics and the dual superconductivity mechanism. Moreover, extracted observables—such as the string tension and the root-mean-square width of the flux tube—exhibit quantitative agreement with lattice QCD benchmarks. These results highlight the potential of ML to streamline the analysis of complex non-perturbative phenomena in QCD, and point to promising avenues for future computational exploration.

\section*{Acknowledgments}
This work has been supported by the National Key R$\&$D Program of China (Grant NO. 2024YFE0109800 and 2024YFE0109802).


\bibliography{refs}

\begin{thebibliography}{63}%
\makeatletter
\providecommand \@ifxundefined [1]{%
 \@ifx{#1\undefined}
}%
\providecommand \@ifnum [1]{%
 \ifnum #1\expandafter \@firstoftwo
 \else \expandafter \@secondoftwo
 \fi
}%
\providecommand \@ifx [1]{%
 \ifx #1\expandafter \@firstoftwo
 \else \expandafter \@secondoftwo
 \fi
}%
\providecommand \natexlab [1]{#1}%
\providecommand \enquote  [1]{``#1''}%
\providecommand \bibnamefont  [1]{#1}%
\providecommand \bibfnamefont [1]{#1}%
\providecommand \citenamefont [1]{#1}%
\providecommand \href@noop [0]{\@secondoftwo}%
\providecommand \href [0]{\begingroup \@sanitize@url \@href}%
\providecommand \@href[1]{\@@startlink{#1}\@@href}%
\providecommand \@@href[1]{\endgroup#1\@@endlink}%
\providecommand \@sanitize@url [0]{\catcode `\\12\catcode `\$12\catcode
  `\&12\catcode `\#12\catcode `\^12\catcode `\_12\catcode `\%12\relax}%
\providecommand \@@startlink[1]{}%
\providecommand \@@endlink[0]{}%
\providecommand \url  [0]{\begingroup\@sanitize@url \@url }%
\providecommand \@url [1]{\endgroup\@href {#1}{\urlprefix }}%
\providecommand \urlprefix  [0]{URL }%
\providecommand \Eprint [0]{\href }%
\providecommand \doibase [0]{https://doi.org/}%
\providecommand \selectlanguage [0]{\@gobble}%
\providecommand \bibinfo  [0]{\@secondoftwo}%
\providecommand \bibfield  [0]{\@secondoftwo}%
\providecommand \translation [1]{[#1]}%
\providecommand \BibitemOpen [0]{}%
\providecommand \bibitemStop [0]{}%
\providecommand \bibitemNoStop [0]{.\EOS\space}%
\providecommand \EOS [0]{\spacefactor3000\relax}%
\providecommand \BibitemShut  [1]{\csname bibitem#1\endcsname}%
\let\auto@bib@innerbib\@empty
\bibitem [{\citenamefont {Nambu}(1974)}]{Nambu:1974zg}%
  \BibitemOpen
  \bibfield  {author} {\bibinfo {author} {\bibfnamefont {Y.}~\bibnamefont
  {Nambu}},\ }\bibfield  {title} {\bibinfo {title} {{Strings, Monopoles and
  Gauge Fields}},\ }\href {https://doi.org/10.1103/PhysRevD.10.4262} {\bibfield
   {journal} {\bibinfo  {journal} {Phys. Rev. D}\ }\textbf {\bibinfo {volume}
  {10}},\ \bibinfo {pages} {4262} (\bibinfo {year} {1974})}\BibitemShut
  {NoStop}%
\bibitem [{\citenamefont {Abrikosov}(1957{\natexlab{a}})}]{Abrikosov:1956sx}%
  \BibitemOpen
  \bibfield  {author} {\bibinfo {author} {\bibfnamefont {A.~A.}\ \bibnamefont
  {Abrikosov}},\ }\bibfield  {title} {\bibinfo {title} {{On the Magnetic
  properties of superconductors of the second group}},\ }\href@noop {}
  {\bibfield  {journal} {\bibinfo  {journal} {Sov. Phys. JETP}\ }\textbf
  {\bibinfo {volume} {5}},\ \bibinfo {pages} {1174} (\bibinfo {year}
  {1957}{\natexlab{a}})}\BibitemShut {NoStop}%
\bibitem [{\citenamefont {Abrikosov}(1957{\natexlab{b}})}]{Abrikosov:1957wnz}%
  \BibitemOpen
  \bibfield  {author} {\bibinfo {author} {\bibfnamefont {A.~A.}\ \bibnamefont
  {Abrikosov}},\ }\bibfield  {title} {\bibinfo {title} {{The magnetic
  properties of superconducting alloys}},\ }\href
  {https://doi.org/10.1016/0022-3697(57)90083-5} {\bibfield  {journal}
  {\bibinfo  {journal} {J. Phys. Chem. Solids}\ }\textbf {\bibinfo {volume}
  {2}},\ \bibinfo {pages} {199} (\bibinfo {year}
  {1957}{\natexlab{b}})}\BibitemShut {NoStop}%
\bibitem [{\citenamefont {Nielsen}\ and\ \citenamefont
  {Olesen}(1973)}]{Nielsen:1973cs}%
  \BibitemOpen
  \bibfield  {author} {\bibinfo {author} {\bibfnamefont {H.~B.}\ \bibnamefont
  {Nielsen}}\ and\ \bibinfo {author} {\bibfnamefont {P.}~\bibnamefont
  {Olesen}},\ }\bibfield  {title} {\bibinfo {title} {{Vortex Line Models for
  Dual Strings}},\ }\href {https://doi.org/10.1016/0550-3213(73)90350-7}
  {\bibfield  {journal} {\bibinfo  {journal} {Nucl. Phys. B}\ }\textbf
  {\bibinfo {volume} {61}},\ \bibinfo {pages} {45} (\bibinfo {year}
  {1973})}\BibitemShut {NoStop}%
\bibitem [{\citenamefont {Kondo}\ \emph {et~al.}(2015)\citenamefont {Kondo},
  \citenamefont {Kato}, \citenamefont {Shibata},\ and\ \citenamefont
  {Shinohara}}]{Kondo:2014sta}%
  \BibitemOpen
  \bibfield  {author} {\bibinfo {author} {\bibfnamefont {K.-I.}\ \bibnamefont
  {Kondo}}, \bibinfo {author} {\bibfnamefont {S.}~\bibnamefont {Kato}},
  \bibinfo {author} {\bibfnamefont {A.}~\bibnamefont {Shibata}},\ and\ \bibinfo
  {author} {\bibfnamefont {T.}~\bibnamefont {Shinohara}},\ }\bibfield  {title}
  {\bibinfo {title} {{Quark confinement: Dual superconductor picture based on a
  non-Abelian Stokes theorem and reformulations of Yang\textendash{}Mills
  theory}},\ }\href {https://doi.org/10.1016/j.physrep.2015.03.002} {\bibfield
  {journal} {\bibinfo  {journal} {Phys. Rept.}\ }\textbf {\bibinfo {volume}
  {579}},\ \bibinfo {pages} {1} (\bibinfo {year} {2015})},\ \Eprint
  {https://arxiv.org/abs/1409.1599} {arXiv:1409.1599 [hep-th]} \BibitemShut
  {NoStop}%
\bibitem [{\citenamefont {Singh}\ \emph {et~al.}(1993)\citenamefont {Singh},
  \citenamefont {Browne},\ and\ \citenamefont {Haymaker}}]{Singh:1993jj}%
  \BibitemOpen
  \bibfield  {author} {\bibinfo {author} {\bibfnamefont {V.}~\bibnamefont
  {Singh}}, \bibinfo {author} {\bibfnamefont {D.~A.}\ \bibnamefont {Browne}},\
  and\ \bibinfo {author} {\bibfnamefont {R.~W.}\ \bibnamefont {Haymaker}},\
  }\bibfield  {title} {\bibinfo {title} {{Structure of Abrikosov vortices in
  SU(2) lattice gauge theory}},\ }\href
  {https://doi.org/10.1016/0370-2693(93)91146-E} {\bibfield  {journal}
  {\bibinfo  {journal} {Phys. Lett. B}\ }\textbf {\bibinfo {volume} {306}},\
  \bibinfo {pages} {115} (\bibinfo {year} {1993})},\ \Eprint
  {https://arxiv.org/abs/hep-lat/9301004} {arXiv:hep-lat/9301004} \BibitemShut
  {NoStop}%
\bibitem [{\citenamefont {Schilling}\ \emph {et~al.}(1999)\citenamefont
  {Schilling}, \citenamefont {Bali},\ and\ \citenamefont
  {Schlichter}}]{Schilling:1998gz}%
  \BibitemOpen
  \bibfield  {author} {\bibinfo {author} {\bibfnamefont {K.}~\bibnamefont
  {Schilling}}, \bibinfo {author} {\bibfnamefont {G.~S.}\ \bibnamefont
  {Bali}},\ and\ \bibinfo {author} {\bibfnamefont {C.}~\bibnamefont
  {Schlichter}},\ }\bibfield  {title} {\bibinfo {title} {{A Ginzburg-Landau
  analysis of the color electric flux tube}},\ }\href
  {https://doi.org/10.1016/S0920-5632(99)85160-3} {\bibfield  {journal}
  {\bibinfo  {journal} {Nucl. Phys. B Proc. Suppl.}\ }\textbf {\bibinfo
  {volume} {73}},\ \bibinfo {pages} {638} (\bibinfo {year} {1999})},\ \Eprint
  {https://arxiv.org/abs/hep-lat/9809039} {arXiv:hep-lat/9809039} \BibitemShut
  {NoStop}%
\bibitem [{\citenamefont {Chernodub}\ \emph {et~al.}(2001)\citenamefont
  {Chernodub}, \citenamefont {Gubarev}, \citenamefont {Polikarpov},\ and\
  \citenamefont {Zakharov}}]{Chernodub:2000rg}%
  \BibitemOpen
  \bibfield  {author} {\bibinfo {author} {\bibfnamefont {M.~N.}\ \bibnamefont
  {Chernodub}}, \bibinfo {author} {\bibfnamefont {F.~V.}\ \bibnamefont
  {Gubarev}}, \bibinfo {author} {\bibfnamefont {M.~I.}\ \bibnamefont
  {Polikarpov}},\ and\ \bibinfo {author} {\bibfnamefont {V.~I.}\ \bibnamefont
  {Zakharov}},\ }\bibfield  {title} {\bibinfo {title} {{Towards Abelian - like
  formulation of the dual gluodynamics}},\ }\href
  {https://doi.org/10.1016/S0550-3213(01)00011-6} {\bibfield  {journal}
  {\bibinfo  {journal} {Nucl. Phys. B}\ }\textbf {\bibinfo {volume} {600}},\
  \bibinfo {pages} {163} (\bibinfo {year} {2001})},\ \Eprint
  {https://arxiv.org/abs/hep-th/0010265} {arXiv:hep-th/0010265} \BibitemShut
  {NoStop}%
\bibitem [{\citenamefont {Chernodub}\ \emph {et~al.}(2005)\citenamefont
  {Chernodub}, \citenamefont {Ishiguro}, \citenamefont {Mori}, \citenamefont
  {Nakamura}, \citenamefont {Polikarpov}, \citenamefont {Sekido}, \citenamefont
  {Suzuki},\ and\ \citenamefont {Zakharov}}]{Chernodub:2005gz}%
  \BibitemOpen
  \bibfield  {author} {\bibinfo {author} {\bibfnamefont {M.~N.}\ \bibnamefont
  {Chernodub}}, \bibinfo {author} {\bibfnamefont {K.}~\bibnamefont {Ishiguro}},
  \bibinfo {author} {\bibfnamefont {Y.}~\bibnamefont {Mori}}, \bibinfo {author}
  {\bibfnamefont {Y.}~\bibnamefont {Nakamura}}, \bibinfo {author}
  {\bibfnamefont {M.~I.}\ \bibnamefont {Polikarpov}}, \bibinfo {author}
  {\bibfnamefont {T.}~\bibnamefont {Sekido}}, \bibinfo {author} {\bibfnamefont
  {T.}~\bibnamefont {Suzuki}},\ and\ \bibinfo {author} {\bibfnamefont {V.~I.}\
  \bibnamefont {Zakharov}},\ }\bibfield  {title} {\bibinfo {title} {{Vacuum
  type of SU(2) gluodynamics in maximally Abelian and Landau gauges}},\ }\href
  {https://doi.org/10.1103/PhysRevD.72.074505} {\bibfield  {journal} {\bibinfo
  {journal} {Phys. Rev. D}\ }\textbf {\bibinfo {volume} {72}},\ \bibinfo
  {pages} {074505} (\bibinfo {year} {2005})},\ \Eprint
  {https://arxiv.org/abs/hep-lat/0508004} {arXiv:hep-lat/0508004} \BibitemShut
  {NoStop}%
\bibitem [{\citenamefont {Cea}\ \emph {et~al.}(2012)\citenamefont {Cea},
  \citenamefont {Cosmai},\ and\ \citenamefont {Papa}}]{Cea:2012qw}%
  \BibitemOpen
  \bibfield  {author} {\bibinfo {author} {\bibfnamefont {P.}~\bibnamefont
  {Cea}}, \bibinfo {author} {\bibfnamefont {L.}~\bibnamefont {Cosmai}},\ and\
  \bibinfo {author} {\bibfnamefont {A.}~\bibnamefont {Papa}},\ }\bibfield
  {title} {\bibinfo {title} {{Chromoelectric flux tubes and coherence length in
  QCD}},\ }\href {https://doi.org/10.1103/PhysRevD.86.054501} {\bibfield
  {journal} {\bibinfo  {journal} {Phys. Rev. D}\ }\textbf {\bibinfo {volume}
  {86}},\ \bibinfo {pages} {054501} (\bibinfo {year} {2012})},\ \Eprint
  {https://arxiv.org/abs/1208.1362} {arXiv:1208.1362 [hep-lat]} \BibitemShut
  {NoStop}%
\bibitem [{\citenamefont {Suzuki}\ \emph {et~al.}(2009)\citenamefont {Suzuki},
  \citenamefont {Hasegawa}, \citenamefont {Ishiguro}, \citenamefont {Koma},\
  and\ \citenamefont {Sekido}}]{Suzuki:2009xy}%
  \BibitemOpen
  \bibfield  {author} {\bibinfo {author} {\bibfnamefont {T.}~\bibnamefont
  {Suzuki}}, \bibinfo {author} {\bibfnamefont {M.}~\bibnamefont {Hasegawa}},
  \bibinfo {author} {\bibfnamefont {K.}~\bibnamefont {Ishiguro}}, \bibinfo
  {author} {\bibfnamefont {Y.}~\bibnamefont {Koma}},\ and\ \bibinfo {author}
  {\bibfnamefont {T.}~\bibnamefont {Sekido}},\ }\bibfield  {title} {\bibinfo
  {title} {{Gauge invariance of color confinement due to the dual Meissner
  effect caused by Abelian monopoles}},\ }\href
  {https://doi.org/10.1103/PhysRevD.80.054504} {\bibfield  {journal} {\bibinfo
  {journal} {Phys. Rev. D}\ }\textbf {\bibinfo {volume} {80}},\ \bibinfo
  {pages} {054504} (\bibinfo {year} {2009})},\ \Eprint
  {https://arxiv.org/abs/0907.0583} {arXiv:0907.0583 [hep-lat]} \BibitemShut
  {NoStop}%
\bibitem [{\citenamefont {Clem}(1975)}]{clem1975simple}%
  \BibitemOpen
  \bibfield  {author} {\bibinfo {author} {\bibfnamefont {J.~R.}\ \bibnamefont
  {Clem}},\ }\bibfield  {title} {\bibinfo {title} {Simple model for the vortex
  core in a type ii superconductor},\ }\href@noop {} {\bibfield  {journal}
  {\bibinfo  {journal} {Journal of Low Temperature Physics}\ }\textbf {\bibinfo
  {volume} {18}},\ \bibinfo {pages} {427} (\bibinfo {year} {1975})}\BibitemShut
  {NoStop}%
\bibitem [{\citenamefont {Larkoski}\ \emph {et~al.}(2020)\citenamefont
  {Larkoski}, \citenamefont {Moult},\ and\ \citenamefont
  {Nachman}}]{Larkoski:2017jix}%
  \BibitemOpen
  \bibfield  {author} {\bibinfo {author} {\bibfnamefont {A.~J.}\ \bibnamefont
  {Larkoski}}, \bibinfo {author} {\bibfnamefont {I.}~\bibnamefont {Moult}},\
  and\ \bibinfo {author} {\bibfnamefont {B.}~\bibnamefont {Nachman}},\
  }\bibfield  {title} {\bibinfo {title} {{Jet Substructure at the Large Hadron
  Collider: A Review of Recent Advances in Theory and Machine Learning}},\
  }\href {https://doi.org/10.1016/j.physrep.2019.11.001} {\bibfield  {journal}
  {\bibinfo  {journal} {Phys. Rept.}\ }\textbf {\bibinfo {volume} {841}},\
  \bibinfo {pages} {1} (\bibinfo {year} {2020})},\ \Eprint
  {https://arxiv.org/abs/1709.04464} {arXiv:1709.04464 [hep-ph]} \BibitemShut
  {NoStop}%
\bibitem [{\citenamefont {Guest}\ \emph {et~al.}(2018)\citenamefont {Guest},
  \citenamefont {Cranmer},\ and\ \citenamefont {Whiteson}}]{Guest:2018yhq}%
  \BibitemOpen
  \bibfield  {author} {\bibinfo {author} {\bibfnamefont {D.}~\bibnamefont
  {Guest}}, \bibinfo {author} {\bibfnamefont {K.}~\bibnamefont {Cranmer}},\
  and\ \bibinfo {author} {\bibfnamefont {D.}~\bibnamefont {Whiteson}},\
  }\bibfield  {title} {\bibinfo {title} {{Deep Learning and its Application to
  LHC Physics}},\ }\href {https://doi.org/10.1146/annurev-nucl-101917-021019}
  {\bibfield  {journal} {\bibinfo  {journal} {Ann. Rev. Nucl. Part. Sci.}\
  }\textbf {\bibinfo {volume} {68}},\ \bibinfo {pages} {161} (\bibinfo {year}
  {2018})},\ \Eprint {https://arxiv.org/abs/1806.11484} {arXiv:1806.11484
  [hep-ex]} \BibitemShut {NoStop}%
\bibitem [{\citenamefont {Radovic}\ \emph {et~al.}(2018)\citenamefont
  {Radovic}, \citenamefont {Williams}, \citenamefont {Rousseau}, \citenamefont
  {Kagan}, \citenamefont {Bonacorsi}, \citenamefont {Himmel}, \citenamefont
  {Aurisano}, \citenamefont {Terao},\ and\ \citenamefont
  {Wongjirad}}]{Radovic:2018dip}%
  \BibitemOpen
  \bibfield  {author} {\bibinfo {author} {\bibfnamefont {A.}~\bibnamefont
  {Radovic}}, \bibinfo {author} {\bibfnamefont {M.}~\bibnamefont {Williams}},
  \bibinfo {author} {\bibfnamefont {D.}~\bibnamefont {Rousseau}}, \bibinfo
  {author} {\bibfnamefont {M.}~\bibnamefont {Kagan}}, \bibinfo {author}
  {\bibfnamefont {D.}~\bibnamefont {Bonacorsi}}, \bibinfo {author}
  {\bibfnamefont {A.}~\bibnamefont {Himmel}}, \bibinfo {author} {\bibfnamefont
  {A.}~\bibnamefont {Aurisano}}, \bibinfo {author} {\bibfnamefont
  {K.}~\bibnamefont {Terao}},\ and\ \bibinfo {author} {\bibfnamefont
  {T.}~\bibnamefont {Wongjirad}},\ }\bibfield  {title} {\bibinfo {title}
  {{Machine learning at the energy and intensity frontiers of particle
  physics}},\ }\href {https://doi.org/10.1038/s41586-018-0361-2} {\bibfield
  {journal} {\bibinfo  {journal} {Nature}\ }\textbf {\bibinfo {volume} {560}},\
  \bibinfo {pages} {41} (\bibinfo {year} {2018})}\BibitemShut {NoStop}%
\bibitem [{\citenamefont {Albertsson}\ \emph {et~al.}(2018)\citenamefont
  {Albertsson} \emph {et~al.}}]{Albertsson:2018maf}%
  \BibitemOpen
  \bibfield  {author} {\bibinfo {author} {\bibfnamefont {K.}~\bibnamefont
  {Albertsson}} \emph {et~al.},\ }\bibfield  {title} {\bibinfo {title}
  {{Machine Learning in High Energy Physics Community White Paper}},\ }\href
  {https://doi.org/10.1088/1742-6596/1085/2/022008} {\bibfield  {journal}
  {\bibinfo  {journal} {J. Phys. Conf. Ser.}\ }\textbf {\bibinfo {volume}
  {1085}},\ \bibinfo {pages} {022008} (\bibinfo {year} {2018})},\ \Eprint
  {https://arxiv.org/abs/1807.02876} {arXiv:1807.02876 [physics.comp-ph]}
  \BibitemShut {NoStop}%
\bibitem [{\citenamefont {Carleo}\ \emph {et~al.}(2019)\citenamefont {Carleo},
  \citenamefont {Cirac}, \citenamefont {Cranmer}, \citenamefont {Daudet},
  \citenamefont {Schuld}, \citenamefont {Tishby}, \citenamefont
  {Vogt-Maranto},\ and\ \citenamefont {Zdeborov\'a}}]{Carleo:2019ptp}%
  \BibitemOpen
  \bibfield  {author} {\bibinfo {author} {\bibfnamefont {G.}~\bibnamefont
  {Carleo}}, \bibinfo {author} {\bibfnamefont {I.}~\bibnamefont {Cirac}},
  \bibinfo {author} {\bibfnamefont {K.}~\bibnamefont {Cranmer}}, \bibinfo
  {author} {\bibfnamefont {L.}~\bibnamefont {Daudet}}, \bibinfo {author}
  {\bibfnamefont {M.}~\bibnamefont {Schuld}}, \bibinfo {author} {\bibfnamefont
  {N.}~\bibnamefont {Tishby}}, \bibinfo {author} {\bibfnamefont
  {L.}~\bibnamefont {Vogt-Maranto}},\ and\ \bibinfo {author} {\bibfnamefont
  {L.}~\bibnamefont {Zdeborov\'a}},\ }\bibfield  {title} {\bibinfo {title}
  {{Machine learning and the physical sciences}},\ }\href
  {https://doi.org/10.1103/RevModPhys.91.045002} {\bibfield  {journal}
  {\bibinfo  {journal} {Rev. Mod. Phys.}\ }\textbf {\bibinfo {volume} {91}},\
  \bibinfo {pages} {045002} (\bibinfo {year} {2019})},\ \Eprint
  {https://arxiv.org/abs/1903.10563} {arXiv:1903.10563 [physics.comp-ph]}
  \BibitemShut {NoStop}%
\bibitem [{\citenamefont {Bourilkov}(2020)}]{Bourilkov:2019yoi}%
  \BibitemOpen
  \bibfield  {author} {\bibinfo {author} {\bibfnamefont {D.}~\bibnamefont
  {Bourilkov}},\ }\bibfield  {title} {\bibinfo {title} {{Machine and Deep
  Learning Applications in Particle Physics}},\ }\href
  {https://doi.org/10.1142/S0217751X19300199} {\bibfield  {journal} {\bibinfo
  {journal} {Int. J. Mod. Phys. A}\ }\textbf {\bibinfo {volume} {34}},\
  \bibinfo {pages} {1930019} (\bibinfo {year} {2020})},\ \Eprint
  {https://arxiv.org/abs/1912.08245} {arXiv:1912.08245 [physics.data-an]}
  \BibitemShut {NoStop}%
\bibitem [{\citenamefont {Schwartz}(2021)}]{Schwartz:2021ftp}%
  \BibitemOpen
  \bibfield  {author} {\bibinfo {author} {\bibfnamefont {M.~D.}\ \bibnamefont
  {Schwartz}},\ }\bibfield  {title} {\bibinfo {title} {{Modern Machine Learning
  and Particle Physics}}\ }\href {https://doi.org/10.1162/99608f92.beeb1183}
  {10.1162/99608f92.beeb1183} (\bibinfo {year} {2021}),\ \Eprint
  {https://arxiv.org/abs/2103.12226} {arXiv:2103.12226 [hep-ph]} \BibitemShut
  {NoStop}%
\bibitem [{\citenamefont {Karagiorgi}\ \emph {et~al.}(2021)\citenamefont
  {Karagiorgi}, \citenamefont {Kasieczka}, \citenamefont {Kravitz},
  \citenamefont {Nachman},\ and\ \citenamefont {Shih}}]{Karagiorgi:2021ngt}%
  \BibitemOpen
  \bibfield  {author} {\bibinfo {author} {\bibfnamefont {G.}~\bibnamefont
  {Karagiorgi}}, \bibinfo {author} {\bibfnamefont {G.}~\bibnamefont
  {Kasieczka}}, \bibinfo {author} {\bibfnamefont {S.}~\bibnamefont {Kravitz}},
  \bibinfo {author} {\bibfnamefont {B.}~\bibnamefont {Nachman}},\ and\ \bibinfo
  {author} {\bibfnamefont {D.}~\bibnamefont {Shih}},\ }\bibfield  {title}
  {\bibinfo {title} {{Machine Learning in the Search for New Fundamental
  Physics}},\ }\href@noop {} {\  (\bibinfo {year} {2021})},\ \Eprint
  {https://arxiv.org/abs/2112.03769} {arXiv:2112.03769 [hep-ph]} \BibitemShut
  {NoStop}%
\bibitem [{\citenamefont {Boehnlein}\ \emph {et~al.}(2022)\citenamefont
  {Boehnlein} \emph {et~al.}}]{Boehnlein:2021eym}%
  \BibitemOpen
  \bibfield  {author} {\bibinfo {author} {\bibfnamefont {A.}~\bibnamefont
  {Boehnlein}} \emph {et~al.},\ }\bibfield  {title} {\bibinfo {title}
  {{Colloquium: Machine learning in nuclear physics}},\ }\href
  {https://doi.org/10.1103/RevModPhys.94.031003} {\bibfield  {journal}
  {\bibinfo  {journal} {Rev. Mod. Phys.}\ }\textbf {\bibinfo {volume} {94}},\
  \bibinfo {pages} {031003} (\bibinfo {year} {2022})},\ \Eprint
  {https://arxiv.org/abs/2112.02309} {arXiv:2112.02309 [nucl-th]} \BibitemShut
  {NoStop}%
\bibitem [{\citenamefont {Shanahan}\ \emph {et~al.}(2022)\citenamefont
  {Shanahan} \emph {et~al.}}]{Shanahan:2022ifi}%
  \BibitemOpen
  \bibfield  {author} {\bibinfo {author} {\bibfnamefont {P.}~\bibnamefont
  {Shanahan}} \emph {et~al.},\ }\bibfield  {title} {\bibinfo {title} {{Snowmass
  2021 Computational Frontier CompF03 Topical Group Report: Machine
  Learning}},\ }\href@noop {} {\  (\bibinfo {year} {2022})},\ \Eprint
  {https://arxiv.org/abs/2209.07559} {arXiv:2209.07559 [physics.comp-ph]}
  \BibitemShut {NoStop}%
\bibitem [{\citenamefont {Yang}\ \emph {et~al.}(2023)\citenamefont {Yang},
  \citenamefont {He}, \citenamefont {Chen}, \citenamefont {Ke}, \citenamefont
  {Pang},\ and\ \citenamefont {Wang}}]{Yang:2022yfr}%
  \BibitemOpen
  \bibfield  {author} {\bibinfo {author} {\bibfnamefont {Z.}~\bibnamefont
  {Yang}}, \bibinfo {author} {\bibfnamefont {Y.}~\bibnamefont {He}}, \bibinfo
  {author} {\bibfnamefont {W.}~\bibnamefont {Chen}}, \bibinfo {author}
  {\bibfnamefont {W.-Y.}\ \bibnamefont {Ke}}, \bibinfo {author} {\bibfnamefont
  {L.-G.}\ \bibnamefont {Pang}},\ and\ \bibinfo {author} {\bibfnamefont
  {X.-N.}\ \bibnamefont {Wang}},\ }\bibfield  {title} {\bibinfo {title} {{Deep
  learning assisted jet tomography for the study of Mach cones in QGP}},\
  }\href {https://doi.org/10.1140/epjc/s10052-023-11807-1} {\bibfield
  {journal} {\bibinfo  {journal} {Eur. Phys. J. C}\ }\textbf {\bibinfo {volume}
  {83}},\ \bibinfo {pages} {652} (\bibinfo {year} {2023})},\ \Eprint
  {https://arxiv.org/abs/2206.02393} {arXiv:2206.02393 [nucl-th]} \BibitemShut
  {NoStop}%
\bibitem [{\citenamefont {Li}\ \emph {et~al.}(2023)\citenamefont {Li},
  \citenamefont {L\"u}, \citenamefont {Pang},\ and\ \citenamefont
  {Qin}}]{Li:2022ozl}%
  \BibitemOpen
  \bibfield  {author} {\bibinfo {author} {\bibfnamefont {F.-P.}\ \bibnamefont
  {Li}}, \bibinfo {author} {\bibfnamefont {H.-L.}\ \bibnamefont {L\"u}},
  \bibinfo {author} {\bibfnamefont {L.-G.}\ \bibnamefont {Pang}},\ and\
  \bibinfo {author} {\bibfnamefont {G.-Y.}\ \bibnamefont {Qin}},\ }\bibfield
  {title} {\bibinfo {title} {{Deep-learning quasi-particle masses from QCD
  equation of state}},\ }\href {https://doi.org/10.1016/j.physletb.2023.138088}
  {\bibfield  {journal} {\bibinfo  {journal} {Phys. Lett. B}\ }\textbf
  {\bibinfo {volume} {844}},\ \bibinfo {pages} {138088} (\bibinfo {year}
  {2023})},\ \Eprint {https://arxiv.org/abs/2211.07994} {arXiv:2211.07994
  [hep-ph]} \BibitemShut {NoStop}%
\bibitem [{\citenamefont {He}\ \emph {et~al.}(2023)\citenamefont {He},
  \citenamefont {Ma}, \citenamefont {Pang}, \citenamefont {Song},\ and\
  \citenamefont {Zhou}}]{He:2023zin}%
  \BibitemOpen
  \bibfield  {author} {\bibinfo {author} {\bibfnamefont {W.-B.}\ \bibnamefont
  {He}}, \bibinfo {author} {\bibfnamefont {Y.-G.}\ \bibnamefont {Ma}}, \bibinfo
  {author} {\bibfnamefont {L.-G.}\ \bibnamefont {Pang}}, \bibinfo {author}
  {\bibfnamefont {H.-C.}\ \bibnamefont {Song}},\ and\ \bibinfo {author}
  {\bibfnamefont {K.}~\bibnamefont {Zhou}},\ }\bibfield  {title} {\bibinfo
  {title} {{High-energy nuclear physics meets machine learning}},\ }\href
  {https://doi.org/10.1007/s41365-023-01233-z} {\bibfield  {journal} {\bibinfo
  {journal} {Nucl. Sci. Tech.}\ }\textbf {\bibinfo {volume} {34}},\ \bibinfo
  {pages} {88} (\bibinfo {year} {2023})},\ \Eprint
  {https://arxiv.org/abs/2303.06752} {arXiv:2303.06752 [hep-ph]} \BibitemShut
  {NoStop}%
\bibitem [{\citenamefont {Zhou}\ \emph {et~al.}(2024)\citenamefont {Zhou},
  \citenamefont {Wang}, \citenamefont {Pang},\ and\ \citenamefont
  {Shi}}]{Zhou:2023pti}%
  \BibitemOpen
  \bibfield  {author} {\bibinfo {author} {\bibfnamefont {K.}~\bibnamefont
  {Zhou}}, \bibinfo {author} {\bibfnamefont {L.}~\bibnamefont {Wang}}, \bibinfo
  {author} {\bibfnamefont {L.-G.}\ \bibnamefont {Pang}},\ and\ \bibinfo
  {author} {\bibfnamefont {S.}~\bibnamefont {Shi}},\ }\bibfield  {title}
  {\bibinfo {title} {{Exploring QCD matter in extreme conditions with Machine
  Learning}},\ }\href {https://doi.org/10.1016/j.ppnp.2023.104084} {\bibfield
  {journal} {\bibinfo  {journal} {Prog. Part. Nucl. Phys.}\ }\textbf {\bibinfo
  {volume} {135}},\ \bibinfo {pages} {104084} (\bibinfo {year} {2024})},\
  \Eprint {https://arxiv.org/abs/2303.15136} {arXiv:2303.15136 [hep-ph]}
  \BibitemShut {NoStop}%
\bibitem [{\citenamefont {Zhou}\ \emph {et~al.}(2023)\citenamefont {Zhou},
  \citenamefont {Pang}, \citenamefont {Shi},\ and\ \citenamefont
  {Stoecker}}]{Zhou:2023tvv}%
  \BibitemOpen
  \bibfield  {author} {\bibinfo {author} {\bibfnamefont {K.}~\bibnamefont
  {Zhou}}, \bibinfo {author} {\bibfnamefont {L.}~\bibnamefont {Pang}}, \bibinfo
  {author} {\bibfnamefont {S.}~\bibnamefont {Shi}},\ and\ \bibinfo {author}
  {\bibfnamefont {H.}~\bibnamefont {Stoecker}},\ }\bibfield  {title} {\bibinfo
  {title} {{Deep Learning for inverse problems in nuclear physics}},\ }\href
  {https://doi.org/10.22323/1.419.0064} {\bibfield  {journal} {\bibinfo
  {journal} {PoS}\ }\textbf {\bibinfo {volume} {FAIRness2022}},\ \bibinfo
  {pages} {064} (\bibinfo {year} {2023})}\BibitemShut {NoStop}%
\bibitem [{\citenamefont {Pang}(2024)}]{Pang:2024kid}%
  \BibitemOpen
  \bibfield  {author} {\bibinfo {author} {\bibfnamefont {L.-G.}\ \bibnamefont
  {Pang}},\ }\bibfield  {title} {\bibinfo {title} {{Studying high-energy
  nuclear physics with machine learning}},\ }\href
  {https://doi.org/10.1142/S0218301324300091} {\bibfield  {journal} {\bibinfo
  {journal} {Int. J. Mod. Phys. E}\ }\textbf {\bibinfo {volume} {33}},\
  \bibinfo {pages} {2430009} (\bibinfo {year} {2024})}\BibitemShut {NoStop}%
\bibitem [{\citenamefont {Ma}\ \emph {et~al.}(2023)\citenamefont {Ma},
  \citenamefont {Pang}, \citenamefont {Wang},\ and\ \citenamefont
  {Zhou}}]{Ma:2023zfj}%
  \BibitemOpen
  \bibfield  {author} {\bibinfo {author} {\bibfnamefont {Y.-G.}\ \bibnamefont
  {Ma}}, \bibinfo {author} {\bibfnamefont {L.-G.}\ \bibnamefont {Pang}},
  \bibinfo {author} {\bibfnamefont {R.}~\bibnamefont {Wang}},\ and\ \bibinfo
  {author} {\bibfnamefont {K.}~\bibnamefont {Zhou}},\ }\bibfield  {title}
  {\bibinfo {title} {{Phase Transition Study Meets Machine Learning}},\ }\href
  {https://doi.org/10.1088/0256-307X/40/12/122101} {\bibfield  {journal}
  {\bibinfo  {journal} {Chin. Phys. Lett.}\ }\textbf {\bibinfo {volume} {40}},\
  \bibinfo {pages} {122101} (\bibinfo {year} {2023})},\ \Eprint
  {https://arxiv.org/abs/2311.07274} {arXiv:2311.07274 [nucl-th]} \BibitemShut
  {NoStop}%
\bibitem [{\citenamefont {Luo}\ \emph {et~al.}(2024)\citenamefont {Luo},
  \citenamefont {Chen}, \citenamefont {Li}, \citenamefont {Li},\ and\
  \citenamefont {Zhou}}]{Luo:2024iwf}%
  \BibitemOpen
  \bibfield  {author} {\bibinfo {author} {\bibfnamefont {O.-Y.}\ \bibnamefont
  {Luo}}, \bibinfo {author} {\bibfnamefont {X.}~\bibnamefont {Chen}}, \bibinfo
  {author} {\bibfnamefont {F.-P.}\ \bibnamefont {Li}}, \bibinfo {author}
  {\bibfnamefont {X.-H.}\ \bibnamefont {Li}},\ and\ \bibinfo {author}
  {\bibfnamefont {K.}~\bibnamefont {Zhou}},\ }\bibfield  {title} {\bibinfo
  {title} {{Neural Network Modeling of Heavy-Quark Potential from
  Holography}},\ }\href@noop {} {\  (\bibinfo {year} {2024})},\ \Eprint
  {https://arxiv.org/abs/2408.03784} {arXiv:2408.03784 [hep-ph]} \BibitemShut
  {NoStop}%
\bibitem [{\citenamefont {Chen}\ \emph {et~al.}(2024)\citenamefont {Chen},
  \citenamefont {Chen}, \citenamefont {Li}, \citenamefont {Zhu},\ and\
  \citenamefont {Zhou}}]{Chen:2024epd}%
  \BibitemOpen
  \bibfield  {author} {\bibinfo {author} {\bibfnamefont {B.}~\bibnamefont
  {Chen}}, \bibinfo {author} {\bibfnamefont {X.}~\bibnamefont {Chen}}, \bibinfo
  {author} {\bibfnamefont {X.}~\bibnamefont {Li}}, \bibinfo {author}
  {\bibfnamefont {Z.-R.}\ \bibnamefont {Zhu}},\ and\ \bibinfo {author}
  {\bibfnamefont {K.}~\bibnamefont {Zhou}},\ }\bibfield  {title} {\bibinfo
  {title} {{Exploring Transport Properties of Quark-Gluon Plasma with a
  Machine-Learning assisted Holographic Approach}},\ }\href@noop {} {\
  (\bibinfo {year} {2024})},\ \Eprint {https://arxiv.org/abs/2404.18217}
  {arXiv:2404.18217 [hep-ph]} \BibitemShut {NoStop}%
\bibitem [{\citenamefont {Omana~Kuttan}\ \emph {et~al.}(2023)\citenamefont
  {Omana~Kuttan}, \citenamefont {Steinheimer}, \citenamefont {Zhou},
  \citenamefont {Redelbach},\ and\ \citenamefont
  {Stoecker}}]{OmanaKuttan:2023bnb}%
  \BibitemOpen
  \bibfield  {author} {\bibinfo {author} {\bibfnamefont {M.}~\bibnamefont
  {Omana~Kuttan}}, \bibinfo {author} {\bibfnamefont {J.}~\bibnamefont
  {Steinheimer}}, \bibinfo {author} {\bibfnamefont {K.}~\bibnamefont {Zhou}},
  \bibinfo {author} {\bibfnamefont {A.}~\bibnamefont {Redelbach}},\ and\
  \bibinfo {author} {\bibfnamefont {H.}~\bibnamefont {Stoecker}},\ }\bibfield
  {title} {\bibinfo {title} {{Extraction of global event features in heavy-ion
  collision experiments using PointNet}},\ }\href
  {https://doi.org/10.22323/1.419.0040} {\bibfield  {journal} {\bibinfo
  {journal} {PoS}\ }\textbf {\bibinfo {volume} {FAIRness2022}},\ \bibinfo
  {pages} {040} (\bibinfo {year} {2023})}\BibitemShut {NoStop}%
\bibitem [{\citenamefont {Shi}\ \emph {et~al.}(2022{\natexlab{a}})\citenamefont
  {Shi}, \citenamefont {Zhou}, \citenamefont {Zhao}, \citenamefont
  {Mukherjee},\ and\ \citenamefont {Zhuang}}]{Shi:2022vfr}%
  \BibitemOpen
  \bibfield  {author} {\bibinfo {author} {\bibfnamefont {S.}~\bibnamefont
  {Shi}}, \bibinfo {author} {\bibfnamefont {K.}~\bibnamefont {Zhou}}, \bibinfo
  {author} {\bibfnamefont {J.}~\bibnamefont {Zhao}}, \bibinfo {author}
  {\bibfnamefont {S.}~\bibnamefont {Mukherjee}},\ and\ \bibinfo {author}
  {\bibfnamefont {P.}~\bibnamefont {Zhuang}},\ }\bibfield  {title} {\bibinfo
  {title} {{From lattice QCD to in-medium heavy-quark interactions via deep
  learning}},\ }\href {https://doi.org/10.22323/1.396.0537} {\bibfield
  {journal} {\bibinfo  {journal} {PoS}\ }\textbf {\bibinfo {volume}
  {LATTICE2021}},\ \bibinfo {pages} {537} (\bibinfo {year}
  {2022}{\natexlab{a}})}\BibitemShut {NoStop}%
\bibitem [{\citenamefont {Shi}\ \emph {et~al.}(2022{\natexlab{b}})\citenamefont
  {Shi}, \citenamefont {Zhou}, \citenamefont {Zhao}, \citenamefont
  {Mukherjee},\ and\ \citenamefont {Zhuang}}]{Shi:2022fei}%
  \BibitemOpen
  \bibfield  {author} {\bibinfo {author} {\bibfnamefont {S.}~\bibnamefont
  {Shi}}, \bibinfo {author} {\bibfnamefont {K.}~\bibnamefont {Zhou}}, \bibinfo
  {author} {\bibfnamefont {J.}~\bibnamefont {Zhao}}, \bibinfo {author}
  {\bibfnamefont {S.}~\bibnamefont {Mukherjee}},\ and\ \bibinfo {author}
  {\bibfnamefont {P.}~\bibnamefont {Zhuang}},\ }\bibfield  {title} {\bibinfo
  {title} {{From lattice QCD to in-medium heavy-quark interactions via deep
  learning}},\ }\href {https://doi.org/10.1051/epjconf/202225904003} {\bibfield
   {journal} {\bibinfo  {journal} {EPJ Web Conf.}\ }\textbf {\bibinfo {volume}
  {259}},\ \bibinfo {pages} {04003} (\bibinfo {year}
  {2022}{\natexlab{b}})}\BibitemShut {NoStop}%
\bibitem [{\citenamefont {Shi}\ \emph {et~al.}(2022{\natexlab{c}})\citenamefont
  {Shi}, \citenamefont {Zhou}, \citenamefont {Zhao}, \citenamefont
  {Mukherjee},\ and\ \citenamefont {Zhuang}}]{Shi:2021qri}%
  \BibitemOpen
  \bibfield  {author} {\bibinfo {author} {\bibfnamefont {S.}~\bibnamefont
  {Shi}}, \bibinfo {author} {\bibfnamefont {K.}~\bibnamefont {Zhou}}, \bibinfo
  {author} {\bibfnamefont {J.}~\bibnamefont {Zhao}}, \bibinfo {author}
  {\bibfnamefont {S.}~\bibnamefont {Mukherjee}},\ and\ \bibinfo {author}
  {\bibfnamefont {P.}~\bibnamefont {Zhuang}},\ }\bibfield  {title} {\bibinfo
  {title} {{Heavy quark potential in the quark-gluon plasma: Deep neural
  network meets lattice quantum chromodynamics}},\ }\href
  {https://doi.org/10.1103/PhysRevD.105.014017} {\bibfield  {journal} {\bibinfo
   {journal} {Phys. Rev. D}\ }\textbf {\bibinfo {volume} {105}},\ \bibinfo
  {pages} {014017} (\bibinfo {year} {2022}{\natexlab{c}})},\ \Eprint
  {https://arxiv.org/abs/2105.07862} {arXiv:2105.07862 [hep-ph]} \BibitemShut
  {NoStop}%
\bibitem [{\citenamefont {Mansouri}\ \emph {et~al.}(2024)\citenamefont
  {Mansouri}, \citenamefont {Bitaghsir~Fadafan},\ and\ \citenamefont
  {Chen}}]{Mansouri:2024uwc}%
  \BibitemOpen
  \bibfield  {author} {\bibinfo {author} {\bibfnamefont {M.}~\bibnamefont
  {Mansouri}}, \bibinfo {author} {\bibfnamefont {K.}~\bibnamefont
  {Bitaghsir~Fadafan}},\ and\ \bibinfo {author} {\bibfnamefont
  {X.}~\bibnamefont {Chen}},\ }\bibfield  {title} {\bibinfo {title}
  {{Holographic complex potential of a quarkonium from deep learning}},\
  }\href@noop {} {\  (\bibinfo {year} {2024})},\ \Eprint
  {https://arxiv.org/abs/2406.06285} {arXiv:2406.06285 [hep-ph]} \BibitemShut
  {NoStop}%
\bibitem [{\citenamefont {Chen}\ and\ \citenamefont
  {Huang}(2024{\natexlab{a}})}]{Chen:2024mmd}%
  \BibitemOpen
  \bibfield  {author} {\bibinfo {author} {\bibfnamefont {X.}~\bibnamefont
  {Chen}}\ and\ \bibinfo {author} {\bibfnamefont {M.}~\bibnamefont {Huang}},\
  }\bibfield  {title} {\bibinfo {title} {{Flavor dependent Critical endpoint
  from holographic QCD through machine learning}},\ }\href@noop {} {\
  (\bibinfo {year} {2024}{\natexlab{a}})},\ \Eprint
  {https://arxiv.org/abs/2405.06179} {arXiv:2405.06179 [hep-ph]} \BibitemShut
  {NoStop}%
\bibitem [{\citenamefont {Chen}\ and\ \citenamefont
  {Huang}(2024{\natexlab{b}})}]{Chen:2024ckb}%
  \BibitemOpen
  \bibfield  {author} {\bibinfo {author} {\bibfnamefont {X.}~\bibnamefont
  {Chen}}\ and\ \bibinfo {author} {\bibfnamefont {M.}~\bibnamefont {Huang}},\
  }\bibfield  {title} {\bibinfo {title} {{Machine learning holographic black
  hole from lattice QCD equation of state}},\ }\href
  {https://doi.org/10.1103/PhysRevD.109.L051902} {\bibfield  {journal}
  {\bibinfo  {journal} {Phys. Rev. D}\ }\textbf {\bibinfo {volume} {109}},\
  \bibinfo {pages} {L051902} (\bibinfo {year} {2024}{\natexlab{b}})},\ \Eprint
  {https://arxiv.org/abs/2401.06417} {arXiv:2401.06417 [hep-ph]} \BibitemShut
  {NoStop}%
\bibitem [{\citenamefont {Wang}\ \emph {et~al.}(2023)\citenamefont {Wang},
  \citenamefont {Dong},\ and\ \citenamefont {Liu}}]{Wang:2023poi}%
  \BibitemOpen
  \bibfield  {author} {\bibinfo {author} {\bibfnamefont {X.-Y.}\ \bibnamefont
  {Wang}}, \bibinfo {author} {\bibfnamefont {C.}~\bibnamefont {Dong}},\ and\
  \bibinfo {author} {\bibfnamefont {X.}~\bibnamefont {Liu}},\ }\bibfield
  {title} {\bibinfo {title} {{Analysis of Strong Coupling Constant with Machine
  Learning and Its Application}},\ }\href
  {https://doi.org/10.1088/0256-307X/41/3/031201} {\bibfield  {journal}
  {\bibinfo  {journal} {Chin. Phys. Lett.}\ }\textbf {\bibinfo {volume} {41}},\
  \bibinfo {pages} {031201} (\bibinfo {year} {2023})},\ \Eprint
  {https://arxiv.org/abs/2304.07682} {arXiv:2304.07682 [hep-ph]} \BibitemShut
  {NoStop}%
\bibitem [{\citenamefont {Bento}\ \emph {et~al.}(2025)\citenamefont {Bento},
  \citenamefont {C\^amara},\ and\ \citenamefont {Seabra}}]{Bento:2025agw}%
  \BibitemOpen
  \bibfield  {author} {\bibinfo {author} {\bibfnamefont {M.~P.}\ \bibnamefont
  {Bento}}, \bibinfo {author} {\bibfnamefont {H.~B.}\ \bibnamefont
  {C\^amara}},\ and\ \bibinfo {author} {\bibfnamefont {J.~F.}\ \bibnamefont
  {Seabra}},\ }\bibfield  {title} {\bibinfo {title} {{Unraveling particle dark
  matter with Physics-Informed Neural Networks}},\ }\href@noop {} {\  (\bibinfo
  {year} {2025})},\ \Eprint {https://arxiv.org/abs/2502.17597}
  {arXiv:2502.17597 [hep-ph]} \BibitemShut {NoStop}%
\bibitem [{\citenamefont {Aarts}\ \emph {et~al.}(2025)\citenamefont {Aarts},
  \citenamefont {Fukushima}, \citenamefont {Hatsuda}, \citenamefont {Ipp},
  \citenamefont {Shi}, \citenamefont {Wang},\ and\ \citenamefont
  {Zhou}}]{Aarts:2025gyp}%
  \BibitemOpen
  \bibfield  {author} {\bibinfo {author} {\bibfnamefont {G.}~\bibnamefont
  {Aarts}}, \bibinfo {author} {\bibfnamefont {K.}~\bibnamefont {Fukushima}},
  \bibinfo {author} {\bibfnamefont {T.}~\bibnamefont {Hatsuda}}, \bibinfo
  {author} {\bibfnamefont {A.}~\bibnamefont {Ipp}}, \bibinfo {author}
  {\bibfnamefont {S.}~\bibnamefont {Shi}}, \bibinfo {author} {\bibfnamefont
  {L.}~\bibnamefont {Wang}},\ and\ \bibinfo {author} {\bibfnamefont
  {K.}~\bibnamefont {Zhou}},\ }\bibfield  {title} {\bibinfo {title}
  {{Physics-driven learning for inverse problems in quantum chromodynamics}},\
  }\href {https://doi.org/10.1038/s42254-024-00798-x} {\bibfield  {journal}
  {\bibinfo  {journal} {Nature Rev. Phys.}\ }\textbf {\bibinfo {volume} {7}},\
  \bibinfo {pages} {154} (\bibinfo {year} {2025})},\ \Eprint
  {https://arxiv.org/abs/2501.05580} {arXiv:2501.05580 [hep-lat]} \BibitemShut
  {NoStop}%
\bibitem [{\citenamefont {Raissi}\ \emph {et~al.}(2019)\citenamefont {Raissi},
  \citenamefont {Perdikaris},\ and\ \citenamefont
  {Karniadakis}}]{raissi2019physics}%
  \BibitemOpen
  \bibfield  {author} {\bibinfo {author} {\bibfnamefont {M.}~\bibnamefont
  {Raissi}}, \bibinfo {author} {\bibfnamefont {P.}~\bibnamefont {Perdikaris}},\
  and\ \bibinfo {author} {\bibfnamefont {G.~E.}\ \bibnamefont {Karniadakis}},\
  }\bibfield  {title} {\bibinfo {title} {Physics-informed neural networks: A
  deep learning framework for solving forward and inverse problems involving
  nonlinear partial differential equations},\ }\href@noop {} {\bibfield
  {journal} {\bibinfo  {journal} {Journal of Computational physics}\ }\textbf
  {\bibinfo {volume} {378}},\ \bibinfo {pages} {686} (\bibinfo {year}
  {2019})}\BibitemShut {NoStop}%
\bibitem [{\citenamefont {Baydin}\ \emph {et~al.}(2015)\citenamefont {Baydin},
  \citenamefont {Pearlmutter}, \citenamefont {Radul},\ and\ \citenamefont
  {Siskind}}]{Baydin:2015tfa}%
  \BibitemOpen
  \bibfield  {author} {\bibinfo {author} {\bibfnamefont {A.~G.}\ \bibnamefont
  {Baydin}}, \bibinfo {author} {\bibfnamefont {B.~A.}\ \bibnamefont
  {Pearlmutter}}, \bibinfo {author} {\bibfnamefont {A.~A.}\ \bibnamefont
  {Radul}},\ and\ \bibinfo {author} {\bibfnamefont {J.~M.}\ \bibnamefont
  {Siskind}},\ }\bibfield  {title} {\bibinfo {title} {{Automatic
  differentiation in machine learning: a survey}},\ }\href@noop {} {\
  (\bibinfo {year} {2015})},\ \Eprint {https://arxiv.org/abs/1502.05767}
  {arXiv:1502.05767 [cs.SC]} \BibitemShut {NoStop}%
\bibitem [{\citenamefont {Amalinadhi}\ \emph {et~al.}(2022)\citenamefont
  {Amalinadhi}, \citenamefont {Palar}, \citenamefont {Stevenson},\ and\
  \citenamefont {Zuhal}}]{amalinadhi2022physics}%
  \BibitemOpen
  \bibfield  {author} {\bibinfo {author} {\bibfnamefont {C.}~\bibnamefont
  {Amalinadhi}}, \bibinfo {author} {\bibfnamefont {P.~S.}\ \bibnamefont
  {Palar}}, \bibinfo {author} {\bibfnamefont {R.}~\bibnamefont {Stevenson}},\
  and\ \bibinfo {author} {\bibfnamefont {L.}~\bibnamefont {Zuhal}},\ }\bibfield
   {title} {\bibinfo {title} {On physics-informed deep learning for solving
  navier-stokes equations},\ }in\ \href@noop {} {\emph {\bibinfo {booktitle}
  {AIAA SCITECH 2022 Forum}}}\ (\bibinfo {year} {2022})\ p.\ \bibinfo {pages}
  {1436}\BibitemShut {NoStop}%
\bibitem [{\citenamefont {Eivazi}\ \emph {et~al.}(2022)\citenamefont {Eivazi},
  \citenamefont {Tahani}, \citenamefont {Schlatter},\ and\ \citenamefont
  {Vinuesa}}]{eivazi2022physics}%
  \BibitemOpen
  \bibfield  {author} {\bibinfo {author} {\bibfnamefont {H.}~\bibnamefont
  {Eivazi}}, \bibinfo {author} {\bibfnamefont {M.}~\bibnamefont {Tahani}},
  \bibinfo {author} {\bibfnamefont {P.}~\bibnamefont {Schlatter}},\ and\
  \bibinfo {author} {\bibfnamefont {R.}~\bibnamefont {Vinuesa}},\ }\bibfield
  {title} {\bibinfo {title} {Physics-informed neural networks for solving
  reynolds-averaged navier--stokes equations},\ }\href@noop {} {\bibfield
  {journal} {\bibinfo  {journal} {Physics of Fluids}\ }\textbf {\bibinfo
  {volume} {34}} (\bibinfo {year} {2022})}\BibitemShut {NoStop}%
\bibitem [{\citenamefont {Baker}\ \emph {et~al.}(2020)\citenamefont {Baker},
  \citenamefont {Cea}, \citenamefont {Chelnokov}, \citenamefont {Cosmai},
  \citenamefont {Cuteri},\ and\ \citenamefont {Papa}}]{Baker:2019gsi}%
  \BibitemOpen
  \bibfield  {author} {\bibinfo {author} {\bibfnamefont {M.}~\bibnamefont
  {Baker}}, \bibinfo {author} {\bibfnamefont {P.}~\bibnamefont {Cea}}, \bibinfo
  {author} {\bibfnamefont {V.}~\bibnamefont {Chelnokov}}, \bibinfo {author}
  {\bibfnamefont {L.}~\bibnamefont {Cosmai}}, \bibinfo {author} {\bibfnamefont
  {F.}~\bibnamefont {Cuteri}},\ and\ \bibinfo {author} {\bibfnamefont
  {A.}~\bibnamefont {Papa}},\ }\bibfield  {title} {\bibinfo {title} {{The
  confining color field in SU(3) gauge theory}},\ }\href
  {https://doi.org/10.1140/epjc/s10052-020-8077-5} {\bibfield  {journal}
  {\bibinfo  {journal} {Eur. Phys. J. C}\ }\textbf {\bibinfo {volume} {80}},\
  \bibinfo {pages} {514} (\bibinfo {year} {2020})},\ \Eprint
  {https://arxiv.org/abs/1912.04739} {arXiv:1912.04739 [hep-lat]} \BibitemShut
  {NoStop}%
\bibitem [{\citenamefont {Lu}\ \emph {et~al.}(2021)\citenamefont {Lu},
  \citenamefont {Meng}, \citenamefont {Mao},\ and\ \citenamefont
  {Karniadakis}}]{lu2021deepxde}%
  \BibitemOpen
  \bibfield  {author} {\bibinfo {author} {\bibfnamefont {L.}~\bibnamefont
  {Lu}}, \bibinfo {author} {\bibfnamefont {X.}~\bibnamefont {Meng}}, \bibinfo
  {author} {\bibfnamefont {Z.}~\bibnamefont {Mao}},\ and\ \bibinfo {author}
  {\bibfnamefont {G.~E.}\ \bibnamefont {Karniadakis}},\ }\bibfield  {title}
  {\bibinfo {title} {{DeepXDE}: A deep learning library for solving
  differential equations},\ }\href {https://doi.org/10.1137/19M1274067}
  {\bibfield  {journal} {\bibinfo  {journal} {SIAM Review}\ }\textbf {\bibinfo
  {volume} {63}},\ \bibinfo {pages} {208} (\bibinfo {year} {2021})}\BibitemShut
  {NoStop}%
\bibitem [{\citenamefont {Cardaci}\ \emph {et~al.}(2011)\citenamefont
  {Cardaci}, \citenamefont {Cea}, \citenamefont {Cosmai}, \citenamefont
  {Falcone},\ and\ \citenamefont {Papa}}]{Cardaci:2010tb}%
  \BibitemOpen
  \bibfield  {author} {\bibinfo {author} {\bibfnamefont {M.~S.}\ \bibnamefont
  {Cardaci}}, \bibinfo {author} {\bibfnamefont {P.}~\bibnamefont {Cea}},
  \bibinfo {author} {\bibfnamefont {L.}~\bibnamefont {Cosmai}}, \bibinfo
  {author} {\bibfnamefont {R.}~\bibnamefont {Falcone}},\ and\ \bibinfo {author}
  {\bibfnamefont {A.}~\bibnamefont {Papa}},\ }\bibfield  {title} {\bibinfo
  {title} {{Chromoelectric flux tubes in QCD}},\ }\href
  {https://doi.org/10.1103/PhysRevD.83.014502} {\bibfield  {journal} {\bibinfo
  {journal} {Phys. Rev. D}\ }\textbf {\bibinfo {volume} {83}},\ \bibinfo
  {pages} {014502} (\bibinfo {year} {2011})},\ \Eprint
  {https://arxiv.org/abs/1011.5803} {arXiv:1011.5803 [hep-lat]} \BibitemShut
  {NoStop}%
\bibitem [{\citenamefont {Cea}\ \emph {et~al.}(2013)\citenamefont {Cea},
  \citenamefont {Cosmai}, \citenamefont {Cuteri},\ and\ \citenamefont
  {Papa}}]{Cea:2013oba}%
  \BibitemOpen
  \bibfield  {author} {\bibinfo {author} {\bibfnamefont {P.}~\bibnamefont
  {Cea}}, \bibinfo {author} {\bibfnamefont {L.}~\bibnamefont {Cosmai}},
  \bibinfo {author} {\bibfnamefont {F.}~\bibnamefont {Cuteri}},\ and\ \bibinfo
  {author} {\bibfnamefont {A.}~\bibnamefont {Papa}},\ }\bibfield  {title}
  {\bibinfo {title} {{Flux tubes and coherence length in the SU(3) vacuum}}\
  }(\bibinfo {year} {2013})\ \Eprint {https://arxiv.org/abs/1310.8423}
  {arXiv:1310.8423 [hep-lat]} \BibitemShut {NoStop}%
\bibitem [{\citenamefont {Cea}\ \emph {et~al.}(2014{\natexlab{a}})\citenamefont
  {Cea}, \citenamefont {Cosmai}, \citenamefont {Cuteri},\ and\ \citenamefont
  {Papa}}]{Cea:2014hma}%
  \BibitemOpen
  \bibfield  {author} {\bibinfo {author} {\bibfnamefont {P.}~\bibnamefont
  {Cea}}, \bibinfo {author} {\bibfnamefont {L.}~\bibnamefont {Cosmai}},
  \bibinfo {author} {\bibfnamefont {F.}~\bibnamefont {Cuteri}},\ and\ \bibinfo
  {author} {\bibfnamefont {A.}~\bibnamefont {Papa}},\ }\bibfield  {title}
  {\bibinfo {title} {{London penetration depth and coherence length of SU(3)
  vacuum flux tubes}},\ }\href {https://doi.org/10.22323/1.214.0350} {\bibfield
   {journal} {\bibinfo  {journal} {PoS}\ }\textbf {\bibinfo {volume}
  {LATTICE2014}},\ \bibinfo {pages} {350} (\bibinfo {year}
  {2014}{\natexlab{a}})},\ \Eprint {https://arxiv.org/abs/1410.4394}
  {arXiv:1410.4394 [hep-lat]} \BibitemShut {NoStop}%
\bibitem [{\citenamefont {Cea}\ \emph {et~al.}(2014{\natexlab{b}})\citenamefont
  {Cea}, \citenamefont {Cosmai}, \citenamefont {Cuteri},\ and\ \citenamefont
  {Papa}}]{Cea:2014uja}%
  \BibitemOpen
  \bibfield  {author} {\bibinfo {author} {\bibfnamefont {P.}~\bibnamefont
  {Cea}}, \bibinfo {author} {\bibfnamefont {L.}~\bibnamefont {Cosmai}},
  \bibinfo {author} {\bibfnamefont {F.}~\bibnamefont {Cuteri}},\ and\ \bibinfo
  {author} {\bibfnamefont {A.}~\bibnamefont {Papa}},\ }\bibfield  {title}
  {\bibinfo {title} {{Flux tubes in the SU(3) vacuum: London penetration depth
  and coherence length}},\ }\href {https://doi.org/10.1103/PhysRevD.89.094505}
  {\bibfield  {journal} {\bibinfo  {journal} {Phys. Rev. D}\ }\textbf {\bibinfo
  {volume} {89}},\ \bibinfo {pages} {094505} (\bibinfo {year}
  {2014}{\natexlab{b}})},\ \Eprint {https://arxiv.org/abs/1404.1172}
  {arXiv:1404.1172 [hep-lat]} \BibitemShut {NoStop}%
\bibitem [{\citenamefont {Baker}\ \emph
  {et~al.}(2025{\natexlab{a}})\citenamefont {Baker}, \citenamefont {Cea},
  \citenamefont {Chelnokov}, \citenamefont {Cosmai},\ and\ \citenamefont
  {Papa}}]{Baker:2024peg}%
  \BibitemOpen
  \bibfield  {author} {\bibinfo {author} {\bibfnamefont {M.}~\bibnamefont
  {Baker}}, \bibinfo {author} {\bibfnamefont {P.}~\bibnamefont {Cea}}, \bibinfo
  {author} {\bibfnamefont {V.}~\bibnamefont {Chelnokov}}, \bibinfo {author}
  {\bibfnamefont {L.}~\bibnamefont {Cosmai}},\ and\ \bibinfo {author}
  {\bibfnamefont {A.}~\bibnamefont {Papa}},\ }\bibfield  {title} {\bibinfo
  {title} {{Unveiling the flux tube structure in full QCD}},\ }\href
  {https://doi.org/10.1140/epjc/s10052-024-13725-2} {\bibfield  {journal}
  {\bibinfo  {journal} {Eur. Phys. J. C}\ }\textbf {\bibinfo {volume} {85}},\
  \bibinfo {pages} {29} (\bibinfo {year} {2025}{\natexlab{a}})},\ \Eprint
  {https://arxiv.org/abs/2409.20168} {arXiv:2409.20168 [hep-lat]} \BibitemShut
  {NoStop}%
\bibitem [{\citenamefont {Baker}\ \emph
  {et~al.}(2025{\natexlab{b}})\citenamefont {Baker}, \citenamefont {Cea},
  \citenamefont {Chelnokov}, \citenamefont {Cosmai},\ and\ \citenamefont
  {Papa}}]{Baker:2024rjq}%
  \BibitemOpen
  \bibfield  {author} {\bibinfo {author} {\bibfnamefont {M.}~\bibnamefont
  {Baker}}, \bibinfo {author} {\bibfnamefont {P.}~\bibnamefont {Cea}}, \bibinfo
  {author} {\bibfnamefont {V.}~\bibnamefont {Chelnokov}}, \bibinfo {author}
  {\bibfnamefont {L.}~\bibnamefont {Cosmai}},\ and\ \bibinfo {author}
  {\bibfnamefont {A.}~\bibnamefont {Papa}},\ }\bibfield  {title} {\bibinfo
  {title} {{Investigating the Flux Tube Structure within Full QCD}},\ }\href
  {https://doi.org/10.22323/1.466.0390} {\bibfield  {journal} {\bibinfo
  {journal} {PoS}\ }\textbf {\bibinfo {volume} {LATTICE2024}},\ \bibinfo
  {pages} {390} (\bibinfo {year} {2025}{\natexlab{b}})},\ \Eprint
  {https://arxiv.org/abs/2411.01886} {arXiv:2411.01886 [hep-lat]} \BibitemShut
  {NoStop}%
\bibitem [{\citenamefont {Kou}\ and\ \citenamefont {Chen}(2025)}]{Kou:2024hzd}%
  \BibitemOpen
  \bibfield  {author} {\bibinfo {author} {\bibfnamefont {W.}~\bibnamefont
  {Kou}}\ and\ \bibinfo {author} {\bibfnamefont {X.}~\bibnamefont {Chen}},\
  }\bibfield  {title} {\bibinfo {title} {{Machine learning insights into
  quark\textendash{}antiquark interactions: probing field distributions and
  string tension in QCD}},\ }\href
  {https://doi.org/10.1140/epjc/s10052-025-13958-9} {\bibfield  {journal}
  {\bibinfo  {journal} {Eur. Phys. J. C}\ }\textbf {\bibinfo {volume} {85}},\
  \bibinfo {pages} {261} (\bibinfo {year} {2025})},\ \Eprint
  {https://arxiv.org/abs/2411.14902} {arXiv:2411.14902 [hep-ph]} \BibitemShut
  {NoStop}%
\bibitem [{\citenamefont {Thuerey}\ \emph {et~al.}(2025)\citenamefont
  {Thuerey}, \citenamefont {Holzschuh}, \citenamefont {Holl}, \citenamefont
  {Kohl}, \citenamefont {Lino}, \citenamefont {Liu}, \citenamefont {Schnell},\
  and\ \citenamefont {Trost}}]{thuerey2025physicsbaseddeeplearning}%
  \BibitemOpen
  \bibfield  {author} {\bibinfo {author} {\bibfnamefont {N.}~\bibnamefont
  {Thuerey}}, \bibinfo {author} {\bibfnamefont {B.}~\bibnamefont {Holzschuh}},
  \bibinfo {author} {\bibfnamefont {P.}~\bibnamefont {Holl}}, \bibinfo {author}
  {\bibfnamefont {G.}~\bibnamefont {Kohl}}, \bibinfo {author} {\bibfnamefont
  {M.}~\bibnamefont {Lino}}, \bibinfo {author} {\bibfnamefont {Q.}~\bibnamefont
  {Liu}}, \bibinfo {author} {\bibfnamefont {P.}~\bibnamefont {Schnell}},\ and\
  \bibinfo {author} {\bibfnamefont {F.}~\bibnamefont {Trost}},\ }\href
  {https://arxiv.org/abs/2109.05237} {\bibinfo {title} {Physics-based deep
  learning}} (\bibinfo {year} {2025}),\ \Eprint
  {https://arxiv.org/abs/2109.05237} {arXiv:2109.05237 [cs.LG]} \BibitemShut
  {NoStop}%
\bibitem [{\citenamefont {Paszke}\ \emph {et~al.}(2017)\citenamefont {Paszke},
  \citenamefont {Gross}, \citenamefont {Chintala}, \citenamefont {Chanan},
  \citenamefont {Yang}, \citenamefont {DeVito}, \citenamefont {Lin},
  \citenamefont {Desmaison}, \citenamefont {Antiga},\ and\ \citenamefont
  {Lerer}}]{paszke2017automatic}%
  \BibitemOpen
  \bibfield  {author} {\bibinfo {author} {\bibfnamefont {A.}~\bibnamefont
  {Paszke}}, \bibinfo {author} {\bibfnamefont {S.}~\bibnamefont {Gross}},
  \bibinfo {author} {\bibfnamefont {S.}~\bibnamefont {Chintala}}, \bibinfo
  {author} {\bibfnamefont {G.}~\bibnamefont {Chanan}}, \bibinfo {author}
  {\bibfnamefont {E.}~\bibnamefont {Yang}}, \bibinfo {author} {\bibfnamefont
  {Z.}~\bibnamefont {DeVito}}, \bibinfo {author} {\bibfnamefont
  {Z.}~\bibnamefont {Lin}}, \bibinfo {author} {\bibfnamefont {A.}~\bibnamefont
  {Desmaison}}, \bibinfo {author} {\bibfnamefont {L.}~\bibnamefont {Antiga}},\
  and\ \bibinfo {author} {\bibfnamefont {A.}~\bibnamefont {Lerer}},\ }\bibfield
   {title} {\bibinfo {title} {Automatic differentiation in pytorch},\
  }\href@noop {} {\  (\bibinfo {year} {2017})}\BibitemShut {NoStop}%
\bibitem [{\citenamefont {Paszke}(2019)}]{paszke2019pytorch}%
  \BibitemOpen
  \bibfield  {author} {\bibinfo {author} {\bibfnamefont {A.}~\bibnamefont
  {Paszke}},\ }\bibfield  {title} {\bibinfo {title} {Pytorch: An imperative
  style, high-performance deep learning library},\ }\href@noop {} {\bibfield
  {journal} {\bibinfo  {journal} {arXiv preprint arXiv:1912.01703}\ } (\bibinfo
  {year} {2019})}\BibitemShut {NoStop}%
\bibitem [{\citenamefont {Abadi}\ \emph {et~al.}(2016)\citenamefont {Abadi},
  \citenamefont {Barham}, \citenamefont {Chen}, \citenamefont {Chen},
  \citenamefont {Davis}, \citenamefont {Dean}, \citenamefont {Devin},
  \citenamefont {Ghemawat}, \citenamefont {Irving}, \citenamefont {Isard} \emph
  {et~al.}}]{abadi2016tensorflow}%
  \BibitemOpen
  \bibfield  {author} {\bibinfo {author} {\bibfnamefont {M.}~\bibnamefont
  {Abadi}}, \bibinfo {author} {\bibfnamefont {P.}~\bibnamefont {Barham}},
  \bibinfo {author} {\bibfnamefont {J.}~\bibnamefont {Chen}}, \bibinfo {author}
  {\bibfnamefont {Z.}~\bibnamefont {Chen}}, \bibinfo {author} {\bibfnamefont
  {A.}~\bibnamefont {Davis}}, \bibinfo {author} {\bibfnamefont
  {J.}~\bibnamefont {Dean}}, \bibinfo {author} {\bibfnamefont {M.}~\bibnamefont
  {Devin}}, \bibinfo {author} {\bibfnamefont {S.}~\bibnamefont {Ghemawat}},
  \bibinfo {author} {\bibfnamefont {G.}~\bibnamefont {Irving}}, \bibinfo
  {author} {\bibfnamefont {M.}~\bibnamefont {Isard}}, \emph {et~al.},\
  }\bibfield  {title} {\bibinfo {title} {$\{$TensorFlow$\}$: a system for
  $\{$Large-Scale$\}$ machine learning},\ }in\ \href@noop {} {\emph {\bibinfo
  {booktitle} {12th USENIX symposium on operating systems design and
  implementation (OSDI 16)}}}\ (\bibinfo {year} {2016})\ pp.\ \bibinfo {pages}
  {265--283}\BibitemShut {NoStop}%
\bibitem [{\citenamefont
  {Lin}(2024)}]{lin2024automaticfunctionaldifferentiationjax}%
  \BibitemOpen
  \bibfield  {author} {\bibinfo {author} {\bibfnamefont {M.}~\bibnamefont
  {Lin}},\ }\href {https://arxiv.org/abs/2311.18727} {\bibinfo {title}
  {Automatic functional differentiation in jax}} (\bibinfo {year} {2024}),\
  \Eprint {https://arxiv.org/abs/2311.18727} {arXiv:2311.18727 [cs.PL]}
  \BibitemShut {NoStop}%
\bibitem [{\citenamefont {{PaddlePaddle
  Contributors}}(2025)}]{paddlepaddle2025}%
  \BibitemOpen
  \bibfield  {author} {\bibinfo {author} {\bibnamefont {{PaddlePaddle
  Contributors}}},\ }\href@noop {} {\bibinfo {title} {Paddlepaddle: An
  open-source deep learning platform}},\ \bibinfo {howpublished}
  {\url{https://github.com/PaddlePaddle/Paddle}} (\bibinfo {year} {2025}),\
  \bibinfo {note} {accessed: 2025-05-22}\BibitemShut {NoStop}%
\bibitem [{\citenamefont {Pang}\ \emph {et~al.}(2019)\citenamefont {Pang},
  \citenamefont {Lu},\ and\ \citenamefont {Karniadakis}}]{Pang_2019}%
  \BibitemOpen
  \bibfield  {author} {\bibinfo {author} {\bibfnamefont {G.}~\bibnamefont
  {Pang}}, \bibinfo {author} {\bibfnamefont {L.}~\bibnamefont {Lu}},\ and\
  \bibinfo {author} {\bibfnamefont {G.~E.}\ \bibnamefont {Karniadakis}},\
  }\bibfield  {title} {\bibinfo {title} {fpinns: Fractional physics-informed
  neural networks},\ }\href {https://doi.org/10.1137/18m1229845} {\bibfield
  {journal} {\bibinfo  {journal} {SIAM Journal on Scientific Computing}\
  }\textbf {\bibinfo {volume} {41}},\ \bibinfo {pages} {A2603–A2626}
  (\bibinfo {year} {2019})}\BibitemShut {NoStop}%
\bibitem [{\citenamefont {Yuan}\ \emph {et~al.}(2022)\citenamefont {Yuan},
  \citenamefont {Ni}, \citenamefont {Deng},\ and\ \citenamefont
  {Hao}}]{yuan2022pinn}%
  \BibitemOpen
  \bibfield  {author} {\bibinfo {author} {\bibfnamefont {L.}~\bibnamefont
  {Yuan}}, \bibinfo {author} {\bibfnamefont {Y.-Q.}\ \bibnamefont {Ni}},
  \bibinfo {author} {\bibfnamefont {X.-Y.}\ \bibnamefont {Deng}},\ and\
  \bibinfo {author} {\bibfnamefont {S.}~\bibnamefont {Hao}},\ }\bibfield
  {title} {\bibinfo {title} {A-pinn: Auxiliary physics informed neural networks
  for forward and inverse problems of nonlinear integro-differential
  equations},\ }\href@noop {} {\bibfield  {journal} {\bibinfo  {journal}
  {Journal of Computational Physics}\ }\textbf {\bibinfo {volume} {462}},\
  \bibinfo {pages} {111260} (\bibinfo {year} {2022})}\BibitemShut {NoStop}%
\bibitem [{\citenamefont {Rohrhofer}\ \emph {et~al.}(2023)\citenamefont
  {Rohrhofer}, \citenamefont {Posch}, \citenamefont {Gößnitzer},\ and\
  \citenamefont {Geiger}}]{IEEE:2023}%
  \BibitemOpen
  \bibfield  {author} {\bibinfo {author} {\bibfnamefont {F.~M.}\ \bibnamefont
  {Rohrhofer}}, \bibinfo {author} {\bibfnamefont {S.}~\bibnamefont {Posch}},
  \bibinfo {author} {\bibfnamefont {C.}~\bibnamefont {Gößnitzer}},\ and\
  \bibinfo {author} {\bibfnamefont {B.~C.}\ \bibnamefont {Geiger}},\ }\bibfield
   {title} {\bibinfo {title} {Data vs. physics: The apparent pareto front of
  physics-informed neural networks},\ }\href
  {https://doi.org/10.1109/ACCESS.2023.3302892} {\bibfield  {journal} {\bibinfo
   {journal} {IEEE Access}\ }\textbf {\bibinfo {volume} {11}},\ \bibinfo
  {pages} {86252} (\bibinfo {year} {2023})}\BibitemShut {NoStop}%
\end{thebibliography}%

\end{document}